\documentclass[
  journal=pasa,
  manuscript=article-type,
  year=2020,
  volume=37,
]{cup-journal}

\usepackage{amsmath}
\usepackage{amssymb}
\usepackage[nopatch]{microtype}
\usepackage{booktabs}
\usepackage{orcidlink}
\usepackage{graphicx}	
\usepackage{placeins}
\usepackage{caption}
\usepackage{hyperref}
\hypersetup{
    colorlinks = true,
    allcolors  = {blue} 
}

\newcommand{\msun}{M$_{\odot}$}
\newcommand{\rsun}{R$_{\odot}$}
\newcommand{\mdot}{M$_\odot$~yr$^{-1}$}
\newcommand{\referee}[1]{#1}

\title[Pre-CE accretion disks]
{Three-dimensional simulations of accretion disks in pre-CE systems}

\author{Ana L. Juarez-Garcia}
\affiliation{School of Mathematical and Physical Sciences, Macquarie University, Balaclava Road, North Ryde, Sydney, NSW 2109, Australia}
\alsoaffiliation{Astrophysics and Space Technologies Research Centre, Macquarie University, Balaclava Road, North Ryde, Sydney, NSW 2109, Australia}
\email[Ana Juarez Garcia]{analourdes.jurezgarca@hdr.mq.edu.au}

\author{Orsola De Marco}
\affiliation{School of Mathematical and Physical Sciences, Macquarie University, Balaclava Road, North Ryde, Sydney, NSW 2109, Australia}
\alsoaffiliation{Astrophysics and Space Technologies Research Centre, Macquarie University, Balaclava Road, North Ryde, Sydney, NSW 2109, Australia}

\author{Fabio De Colle}
\affiliation{Instituto de Ciencias Nucleares, Universidad Nacional Aut{\'o}noma de M{\'e}xico, A. P. 70-543 04510 D. F. Mexico}

\author{Diego L{\'o}pez-C{\'a}mara}
\affiliation{Instituto de Ciencias Nucleares, Universidad Nacional Aut{\'o}noma de M{\'e}xico, A. P. 70-543 04510 D. F. Mexico}
\alsoaffiliation{Investigador por M{\'e}xico, CONAHCYT – Universidad Nacional Aut{\'o}noma de M{\'e}xico, Instituto de Astronom{\'i}a, AP 70-264, CDMX 04510, M{\'e}xico}

\author{Enrique Moreno M{\'e}ndez}
\affiliation{Facultad de Ciencias, Universidad Nacional Aut{\'o}noma de M{\'e}xico, A. P. 70-543 04510 D. F. Mexico}

\author{Jes{\'u}s Carrillo-Santamar{\'i}a}
\affiliation{Instituto de Astronom{\'i}a, Universidad Nacional Aut{\'o}noma de M{\'e}xico, A. P. 70-264 04510 D. F. M\'exico}

\author{Mark Wardle}
\affiliation{School of Mathematical and Physical Sciences, Macquarie University, Balaclava Road, North Ryde, Sydney, NSW 2109, Australia}
\alsoaffiliation{Astrophysics and Space Technologies Research Centre, Macquarie University, Balaclava Road, North Ryde, Sydney, NSW 2109, Australia}


\begin{document}

\begin{abstract}
Before a binary system enters into a common envelope (CE) phase, accretion from the primary star onto the companion star through Roche Lobe overflow (RLOF) will lead to the formation of an accretion disk, which may generate jets. Accretion before and during the CE may alter the outcome of the interaction. Previous studies have considered different aspects of this physical mechanism. Here we study the properties of an accretion disk formed via 3D hydrodynamic simulations of the RLOF mass transfer between  a 7~\msun, red supergiant star and a 1.4~\msun, neutron star companion. We simulate only the volume around the companion for improved resolution. We use a 1D implicit {\sc mesa} simulation of the evolution of the system during 30\,000 years between the on-set of the RLOF and the CE to guide the binary parameters and the mass-transfer rate, while we simulate only 21 years of the last part of the RLOF in 3D using an ideal gas quasi-isothermal equation of state. We expect that a pre-CE disk under these parameters will have a mass of $\sim 5\times 10^{-3}$~\msun\ and a radius of $\sim$40~\rsun\ with a scale height of $\sim$5~\rsun. The temperature profile of the disk is shallower than that predicted by the formalism of Shakura and Sunyaev, but more reasonable cooling physics would need to be included. We stress test these results with respect to a number of physical and numerical parameters, as well as simulation choices, and we expect them to be reasonable within a factor of a few for the mass and 15\% for the radius. We also contextualise our results within those presented in the literature, in particular with respect to the dimensionality of simulations and the adiabatic index. 
\referee{We discuss the measured accretion rate in the context of the Shakura and Sunyaev formalism and debate the viscous mechanisms at play, finishing with a list of prospects for future work. }

\end{abstract}

\section{Introduction}
\label{sec:introduction}

Massive stars ($M \gtrsim 8$~\msun) are almost always in binary and multiple systems with $\sim$70~\% of them involved in various forms of interaction, including tidal interactions and mass transfer, leading eventually to close binaries and mergers \citep[][]{moedisteffano2017}. These interacting binary systems can give rise to high-energy phenomena, such as cataclysm variables \citep[e.g.,][]{warner_1995}, type Ia supernovae \citep[e.g.,][]{ibenandttukov1984, chevalier2012common}, short and long gamma-ray bursts \citep[e.g.,][]{fryer1998helium, Brown2007GRB, ramirez2009maybe}, and gravitational wave emission \citep[e.g.,][]{abbott2016gw151226}.

For a certain range of binary and stellar parameters, the massive binary becomes a high-mass X-ray binary, where a red giant or red supergiant (RGS) feeds mass through the inner Lagrangian point, $L_1$, to a neutron star (or sometimes a black hole) companion in a phase of wind accretion and, possibly, Roche lobe overflow (RLOF). These systems form an accretion disk around the compact companion, which is X-ray bright and may develop jets. If the mass ratio is large, as is the case if the compact object is a neutron star, mass transfer reduces the orbital separation. In addition, the RGSs expand upon loss of mass, further accelerating the mass transfer. For neutron star - RGS systems, it is likely that unstable mass transfer and a common envelope (CE) phase may result \citep{ivanova2013common,tauris2017}. The outcome of this phase depends on whether or not the CE can be ejected before the neutron star merges with the helium core of the RGS. 

Most observed X-ray binaries are undergoing a long-lived phase of stable wind accretion with timescales that depend on the parameters of the system. Some systems may be undergoing RLOF, in which case the evolution timescales are likely much shorter with a possibility of unstable mass transfer and CE (e.g., high mass X-ray binaries may remain in the RLOF state for only 10\,000~yr; \citealt{Savonije77}, see also \citealt{tauris2017}). Although, presumably, X-ray binaries in the stable wind accretion phase are more frequently observed (e.g., Cygnus X-1), it is possible that some may be caught in the faster phase of unstable mass transfer.   \citet{Dickson2024} presented a model of X-ray binary M\,33~X-7, that they believe to be in an unstable mass transfer phase based on a measurement by \citet{Ramachandra22} of the donor substantially overfilling its Roche lobe. 

One open question in the study of massive CE interactions is the impact of the energetic feedback due to accretion of envelope gas onto the compact companion, before and during the in-spiral in the CE. For neutron stars and black hole companions, this could be critically important for the outcome of the interaction. \citet{Shiber2019} simulated an {\it ad hoc} jet, emanating from a companion in a low mass CE interaction to conclude that the jet would aid in unbinding envelope mass with consequences for the  binary separation of the post-CE binary. On the other hand \citet{LopezCamara2022}, using a self-regulated jet, powered by a fraction of the mass accretion rate that reached the inner boundary, showed that the jet would likely be quenched, even if it existed before the companion entered the CE. 

A disk that forms during wind accretion and RLOF could survive inside the CE, be destroyed, or be destroyed and reform, as a result of accretion of envelope material on the secondary star \cite[][]{macleod2015asymmetric, macleod2017common,moreno2017, chamandy2018,lopez2019, Shiber2019,lopez2020disc, mm2022}. So far previous studies have found that the formation of the accretion disk inside a CE depends on the thermal properties of the envelope (adiabatic index of $\lesssim 1.2$). \cite{Murguia-Berthier2017AccretionFormation} pointed out that this phase is only a transitory phase, due to the lack of stellar regions (zones of partial ionisation where $\gamma$ is small enough) where the envelope is compressible enough to form a disk.

In this work, we study the formation of the disk around a 1.4~\msun\ neutron star, caused by RLOF mass transfer from a 7~\msun\ RGS, undergoing unstable mass transfer. We attempt to gain a quantitative idea of the parameters of an accretion disk to, eventually, determine its fate inside the CE.
This work is also intended to contribute to the literature by studying accretion disks in 3D hydrodynamics. The goal is to determine when and how accretion disks form in response to mass accretion through $L_1$ and as a function of a number of physical and numerical parameters to set this study in the broader context of disk formation \citep[e.g.,][]{makita2000two}.

This paper is structured as follows. In Section~\ref{sec:methods} we outline the overall methodology with Sections~\ref{ssec:hydrocode} and \ref{sec:initialconditions} presenting the governing equations and simulation parameters. In Section~\ref{sec:results} we give details of the formation and evolution of the disk and, in Section~\ref{sec:isadisk} the disk parameters. In Section~\ref{sec:physicalandnumericalparam} we discuss the sensitivity of our results to some physical and numerical parameters, while in Section~\ref{sec:Discussion} we present our conclusions.

\section{Methods}
\label{sec:methods}

\begin{figure}
    \centering
   \includegraphics[width=\columnwidth]{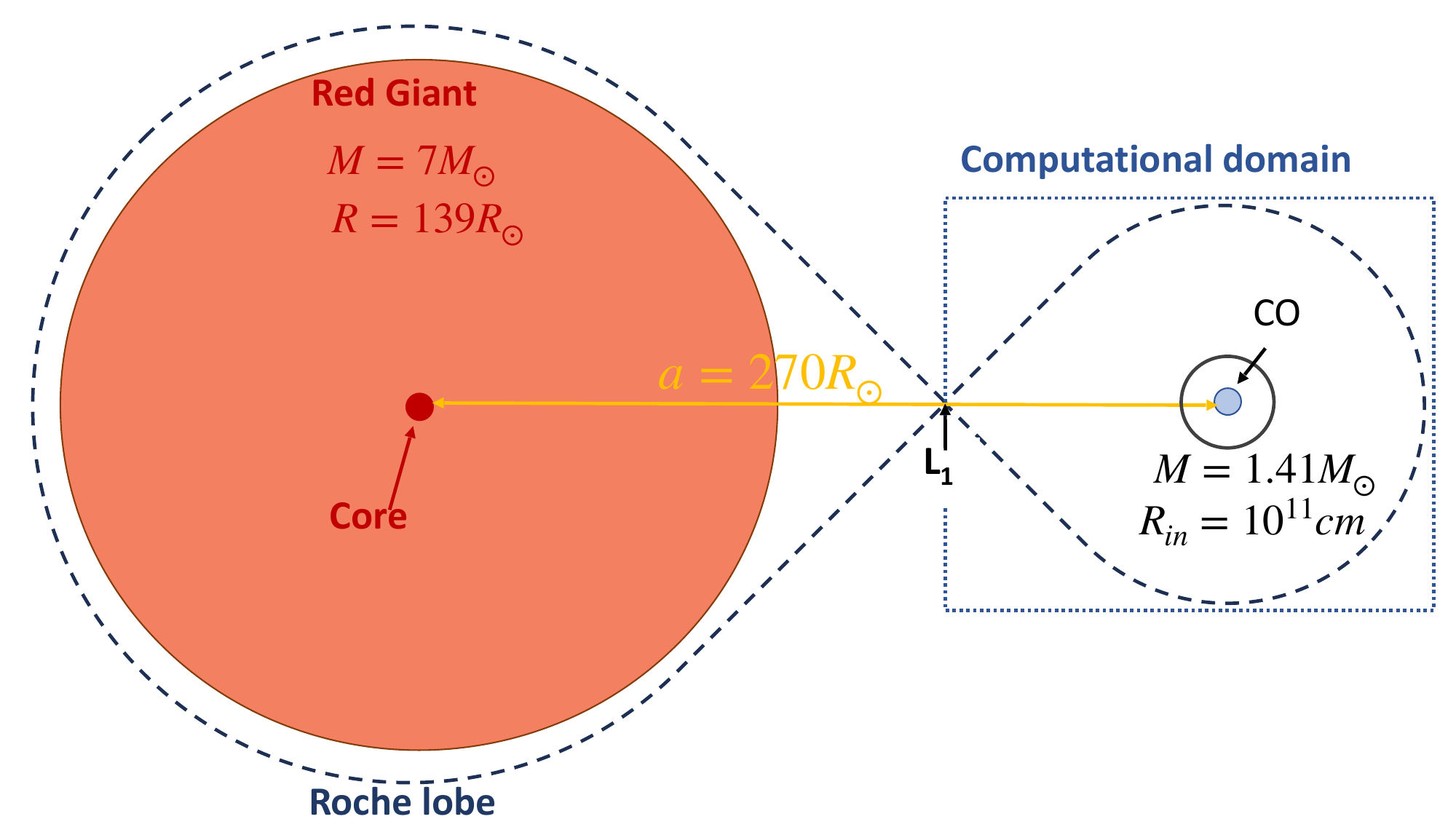}
    \caption{Setup cartoon. The donor star is a red super giant of 7~\msun\ with a radius of 139~\rsun, orbiting a compact object of 1.41~\msun, with a separation of 270~\rsun. The dotted line (centred on the compact object) represents the computational domain in our 3D simulations.}
   \label{fig:cartoon}
\end{figure}

To study the mass transfer phase through the $L_1$ point, we consider a binary system consisting a 7~\msun\ red supergiant as the donor star and a 1.4~\msun\ neutron star as the companion star (see Figure~\ref{fig:cartoon} for a cartoon of the setup). We model the evolution of this binary system, between the RLOF phase and the CE phase, using the 1D implicit code {\sc mesa} (Modules for Experiments in Stellar Astrophysics; version r21.12.1;  
\citealt{paxton2011, paxton2013, paxton2015, paxton2018, paxton2019}). 

We then use use the 3D hydrodynamic numerical code {\sc Mezcal} \citep{decolle2012} to simulate the formation of the accretion disk around the companion star using the mass transfer rate given by the {\sc mesa} model as boundary condition to inject material into the computational domain. The 3D computational domain is represented by a dotted line in Figure~\ref{fig:cartoon}.

To only simulate the region around the accreting star, the 3D simulation is performed in a co-rotating system of reference, centred on the companion which is represented by a point mass particle encircled by an inflow boundary of radius R$_{\rm in}$.  Mass is injected through a nozzle that represents the $L_1$ point, located at the centre of the left boundary face. Below  we motivate the method and the setup and explain the specific assumptions.

\subsection{The hydrodynamic code and its governing equations}
\label{ssec:hydrocode}

To study the formation and stability of accretion disks during the RLOF phase, we run a series of 3D numerical simulations employing the adaptive mesh refinement code {\sc Mezcal} \citep{decolle2012}. The code integrates the hydrodynamic equations in a rotating frame. Self gravity is not included in these simulations.

We solve the three-dimensional Euler equations for an inviscid gas (see, e.g., \citealt{makita2000two}), that is,

\begin{equation}
    \frac{\partial \rho}{\partial t} + \nabla \cdot \left( \rho \textbf{v} \right) = 0\;,
    \label{eq:euler1}
\end{equation}

\begin{equation}
    \frac{\partial \left( \rho \textbf{v} \right)}{\partial t} + \nabla \cdot \left(\rho \textbf{v}\textbf{v} + P I \right) = -\rho \textbf{f}\;,
    \label{eq:euler2}
\end{equation}

\begin{equation}
    \frac{\partial e}{\partial t} + \nabla \cdot \left[\left( e+p \right)\textbf{v} \right] = -\rho \textbf{v}\cdot \textbf{f} \;,
    \label{eq:euler3}
\end{equation}

\noindent representing the evolution of mass density, $\rho$, gas momentum, $\rho \textbf{v}$, and gas energy density, $e$ \referee{(for additional details see \citet{Pringle2014})}. The variable $P$ represents the pressure, $I$ is the identity tensor, $\textbf{v}$ is the velocity vector, and $\textbf{f}$ is the specific force vector. The specific force vector $\textbf{f}~=~\left(f_{\rm x},f_{\rm y},f_{\rm z}\right)$ represents the  gravity forces and fictitious forces associated with the rotating frame.

The binary system consists of a donor star of mass $M_{\rm 1}$, and an accreting star of mass $M_{\rm2}$, with a mass ratio defined as $q~=~M_{\rm 1}/M_{\rm2}$ during their mass transfer phase. Our simulations are conducted in Cartesian coordinates in three dimensions with different levels of resolution. The binary system has an orbital separation $a$, with an orbital frequency $\Omega=(G(M_{\rm 1}+M_{\rm 2})/a^3)^{1/2}$. Then, the orbital period is given by $P=2\pi/\Omega$. The origin of the rotating frame is the accreting star; the donor star is positioned on the left of the accretor (donor position is ($-a$,0,0)). We perform the simulation in dimensionless units using $a$ as the length scale and $a\Omega/2\pi$ as the velocity scale. The time unit is the initial period of the binary. This means that quantities can be rescaled based on these units to physical units. In Table~\ref{tab:factors} we list the scaling factors between code and physical units.

\begin{table}[hbt!]                          
\centering    
\begin{tabular}{l l l}                     
\hline                                          
\hline
 & Code unit& Value in cgs \\
\hline
Length                        & $a$         & $1.85\times10^{13}$  cm                       \\ 
Velocity                      & $a\Omega/2\pi$   & $7.75\times 10^{6}$ cm s$^{-1}$               \\
Mass                          & $M_{\rm tot}$& $1.67\times10^{34}$ g                         \\
Time                          & 2$\pi/\Omega$ & $0.48$ yr                                       \\
Mass transfer rate            & $M_{\rm tot} \Omega/2\pi$  &$6.97\times10^{27}$ g s$^{-1}$               \\
Density                       & $M_{\rm tot}/a^3$    & $2.62\times10^{-6}$ g cm$^{-3}$                    \\
Pressure                      & $M_{\rm tot} \Omega^2  /4\pi^2 a$  & $3.89 \times 10^{8}$ dyne~cm$^{-2}$                 \\
\hline
\end{tabular}
\caption[Rescaling factors...]{Description of units, rescaling factors to physical systems, in terms of the binary separation ($a$), orbital frequency ($\Omega$), and total mass (M$_{\rm tot}$~=~M$_{\rm{don}}$~+~M$_{\rm{acc}}$, where M$_{\rm{don}}$ is the mass of the donor star and  M$_{\rm{acc}}$ is the mass of the companion star).}
\label{tab:factors}                              
\end{table}

With these assumptions, we can now write out the force vector in dimensionless form at each point in the computational domain, remembering that there is no self-gravity, but that we are operating in the rotating frame:
 
\begin{equation}
    f_{\rm x} = - 2v_{\rm y} - \left(x + \frac{q}{1+q}\right) + \frac{1/(1+q)}{|\textbf{r}_1|^3}(x - a) + \frac{q/(1+q)}{|\textbf{r}_2|^3}x,
    \label{eq:xforce}
\end{equation}

\begin{equation}
    f_{\rm y} = 2v_x - y + \frac{1/(1+q)}{|\textbf{r}_1|^3}y + \frac{q/(1+q)}{|\textbf{r}_2|^3}y,
    \label{eq:yforce}
\end{equation}

\begin{equation}
    f_{\rm z} =\frac{1/(1+q)}{|\textbf{r}_1|^3}z + \frac{q/(1+q)}{|\textbf{r}_2|^3}z,
    \label{eq:zforce}
\end{equation}
where $\textbf{r}_1$ and $\textbf{r}_2$ are the distances from the point considered $\textbf{r}~=~\left(x,y,z\right)$ to the centre of each star. In Equations (\ref{eq:xforce}) and (\ref{eq:yforce}), on the right-hand side, the first term represents the Coriolis force and the second term represents the centrifugal force. The last two terms in Equations (\ref{eq:xforce}) and (\ref{eq:yforce}), as well as the terms in Equation (\ref{eq:zforce}), represent the gravity force of each star.

\subsection{Initial conditions}
\label{sec:initialconditions}

\subsubsection{Calculation of the mass transfer rate through \texorpdfstring{L$_{\rm{1}}$}{L1} using a 1D implicit code}
\label{Sec:Mesa}

To determine the mass transfer rate in a binary system between a massive red supergiant donor star and a compact accretor, we use {\sc mesa} to simulate the  evolution of a binary system comprising a 7~\msun, solar metallicity main sequence star with a radius of 57~\rsun\ and a 1.4~\msun\  point mass companion, initially located at an orbital separation of a~=~270~\rsun. The Roche lobe radius of the primary star (R$_{L_1}$) is calculated according to the prescription by \citet{eggleton1983}:
\begin{equation}
    \frac{R_{L_1}}{a} = \frac{0.49 q^{2/3}}{0.6 q ^{2/3}+ln{\left(1+q^{1/3}\right)}},
\end{equation}

\noindent where $a$ is the orbital separation, $q = M_1/M_2$ is the mass ratio between the donor star ($M_1$) and the accreting star ($M_2$). 

\begin{figure}[hbt!]
    \centering
    \includegraphics[width=\columnwidth]{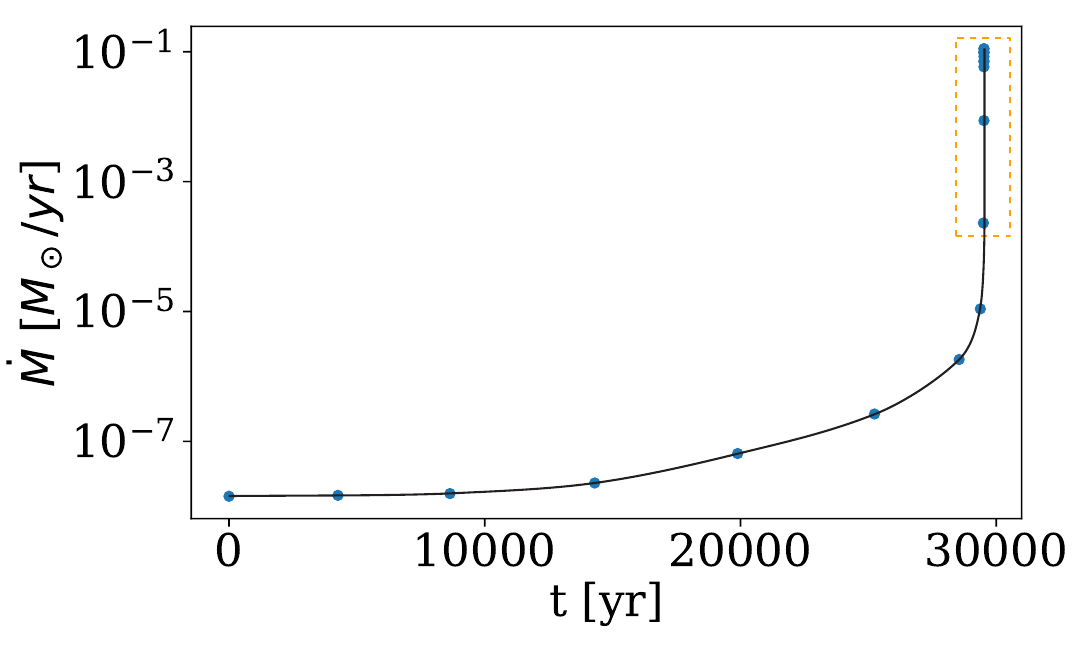}
    \caption{Temporal evolution of the mass transfer rate from the primary to the secondary star in our long-term, {\sc mesa} simulation of the binary system. The initial masses of the stars are 7~\msun\ and 1.4~\msun; they have an initial orbital separation of 270~\rsun\ and an orbital period of 177~days. We indicate with the orange box the mass transfer evolution that we are simulating in 3D.}
    \label{fig:mesamdot}
\end{figure}

We let the {\sc mesa} simulation run for $\sim 4.0 \times 10^{7}$~yrs until the primary expands to fill its Roche lobe and starts to transfer mass to the companion. At this time, called time zero in Figure~\ref{fig:mesamdot}, the red supergiant has a helium-burning core surrounded by a shell that experiences hydrogen burning (primarily through the CNO cycle), along with a massive convective hydrogen envelope. The orbital separation is 270.4~\rsun, 
the red supergiant has a mass of 6.98~\msun, a radius of
134~\rsun, an effective temperature of $3\,981$~K, and a luminosity of $3\,645$~L$_\odot$. The companion star has a mass of 1.4~M$_\odot$, and the mass transfer rate at time zero is $1.4\times 10^{-8}$~\mdot.

We then continue the {\sc mesa} simulation for an additional $\sim$30\,000~yr, during which the mass transfer rate increases to a maximum value of 9.7$\times10^{-2}$~\mdot\ (as we can see in Figure~\ref{fig:mesamdot}). After 30\,000~yrs the mass of the red supergiant is 6.88~\msun\, its radius is 143~\rsun\ and its effective temperature is $\sim$~5\,012~K, while the companion star mass has increased to 1.49~\msun, having accreted some mass. The orbital separation is 243~\rsun. After this point we assume that the mass transfer leads to a CE in a short timescale.
See Table~\ref{tab:MESA parameters} for a summary of the initial and final parameters of the {\sc mesa} simulation. 

\begin{table}
    \centering
    \begin{tabular}{lcc} 
    
    \hline     
    \hline
         Quantity & Initial value & Final value \\
    \hline     
         Donor star &  7.0~\msun & 6.88~\msun\\
         Accretor star &1.4~\msun & 1.49~\msun\\ 
         Orbital separation&  270~\rsun & 243~\rsun  \\ 
         Mass transfer rate & 1.6$\times10^{-9}$~\mdot & 9.7$\times 10^{-2}$~\mdot\\
   \hline
    \end{tabular}
    \caption{Initial (start of RLOF) and final (the time of CE) binary parameters for 30\,000~yr of Roche lobe mass transfer, modeled with the {\sc mesa} code.}
    \label{tab:MESA parameters}
\end{table}

The 3D simulation described next (Section~\ref{sssec:3D}) instead spans only 21 years. This period of time is taken between 29\,500~yr and 29\,521~yr of the 30\,000~yr stretch of the 1D simulation. Hence, the mass transfer rate in the 3D simulation, is prescribed from the 1D simulation to be between 2.3$\times10^{-4}$~\mdot\ and 9.7$\times10^{-2}$~\mdot. In Section~\ref{sec:masstransfer} we will explore the impact that this mass-transfer rate choice has on the the accretion disk parameters. 

\subsubsection{Calculation of disk formation using a 3D explicit code}
\label{sssec:3D}

In the 3D simulations, a 1.4~\msun\ 
point mass particle represents the neutron star companion, at the centre of the domain.  We assume that the donor star, with a mass of 7.0~\msun, is located outside the computational domain at a position $(-a,0,0)$ from the companion star, where $a= 266$~\rsun\ is the binary separation at the start of the hydrodynamic simulation. In this way, the position of $L_{\rm1}$ is at 83.1~\rsun\ ($5.79\times 10^{12}$~cm) from the companion star. 
The gas is injected through the $L_{\rm1}$ point, represented in the simulations by a small rectangular boundary with a variable size given by the cross section of the mass transfer stream, as explained in the following. The setup scheme is shown in Figure~\ref{fig:cartoon}.

The point-mass companion sits inside a spherical inflow boundary of radius R$_{\rm in}$~=1.3~\rsun\ ($9.3\times10^{10}$~cm). \referee{Since cells cannot be empty, we set} the density of gas within this inner boundary to  $\rho_{\rm in}$~=~1.0$\times10^{-20}$~g~cm$^{-3}$ and the pressure to 1.0$\times10^{-8}$~dyne~cm$^{-2}$. \referee{The cells inside the inner boundary are rewritten at every time step to have these values. Outside of the inner boundary} we set a low background density $\rho_{\rm bg}$~=~2.6$\times 10^{-22}$~g~cm$^{-3}$, a temperature T$_{\rm bg}$~=~1.0$\times 10^{5}$ K, and a pressure $P_{\rm bg}~=~8.9\times 10^{-11}$~dyne~cm$^{-2}$. \referee{These initial conditions were chosen to have the lowest density possible before the simulation stops being able to calculate gas advection.} The sensitivity of our simulations to different values of the background conditions is tested in Section~\ref{sec:medium}. The gas adiabatic index is fixed to $\gamma=1.1$ during the whole simulation, based on  \citet{makita2000two}, \citet{macleod2015asymmetric} and \citet{Murguia-Berthier2017AccretionFormation}. See Section~\ref{sec:makita} for further discussion about the adiabatic index in simulations of accretion disks.

Following the \cite{Jackson2017AStars} prescription for optically thin mass transfer, we define an elliptical area centred on L$_{\rm{1}}$,  where most of the material escapes L$_{\rm{1}}$, referred from now on as the ``nozzle". 
The nozzle area is defined by $S_{\rm L_1}= \pi \Delta z \Delta y$, where the two dimensions are related to the pressure scale height in the y- and z-direction, and vary as:
\begin{equation}
    \Delta y = \frac{\sqrt{2}c_T}{\Omega\sqrt{A-1}},  \;\;\; \Delta z = \frac{\sqrt{2}c_T}{\Omega\sqrt{A}},
    \label{eq:nozzle}
\end{equation}
where $c_{\rm T} = \sqrt{\gamma*k_{\rm B} T / \mu}$ is the isothermal sound speed, $k_{\rm B}$ is the Boltzmann constant, $\mu$ is the mean mass of a gas particle, $\Omega$ is the orbital frequency and A is a dimensionless coefficient that depends on the mass ratio, $q$, (equivalently $M_2/M_1$ or $M_1/M_2$ in this equation), defined as
\begin{equation}
    A(q) = 4 + \frac{4.16}{-0.96 + q^{1/3} + q^{-1/3}}.
\end{equation}

We set the length of the nozzle on the $x$-axis to be two cells thick at the coarsest level of refinement, or $\Delta x$~=~1.30~\rsun\ ($9.05\times 10^{10}$~cm). The initial mass injection rate is $\dot{M}_{\rm{ini}}=2.31\times10^{-4}$~\mdot, and it has a subsonic velocity in the $x$ direction with value  $v_{\rm L_1}=77.7\times 10 ^{3}$~cm~s$^{-1}$ \citep[][]{lubow1975gas,Jackson2017AStars,Cehula2023}. The mass injection rate ($\dot{M}$) is interpolated at each time step ($\Delta t$), using the values given by the {\sc mesa} simulation (see Section~\ref{Sec:Mesa}). Using the interpolated values, we calculate the nozzle volume (V$_{\rm{L_1}}$), nozzle density as $\rho_{\rm L_1}~=~\dot{M}~{\Delta t}/V_{\rm L_1}$, and the pressure of the nozzle as $P_{\rm L_1}~=~\rho_{\rm L_1}c_{\rm T}^2$, using the effective temperature of the donor star to calculate $c_{\rm T}$.  

\begin{table*}
    \caption{Simulations' summary: inputs are varied to understand the resilience of the accretion disk parameters to numerical and physical input parameter changes. We vary the initial mass transfer rate (sim-dot-$\#$), injection velocity through L$_{\rm{1}}$ (sim-vel-$\#$), and background temperature and density (sim-bgT-$\#$). We compare each simulation to the reference simulation (sim-0). The majority of the simulations were carried out for 21~yr and start with a donor star of 7~\msun, a companion star of 1.4~\msun, an orbital separation of 266.34~\rsun, and $\Omega$ of 4.18$\times10^{-7}$~s$^{-1}$.
    }
    \centering
    \begin{tabular}{lcccccc}
    \hline
    \hline
         Model&  $\dot{M}_{\rm ini}$
         &  $\dot{M}_{\rm fin}$&  $\rho_{\rm bg}$ & $P_{\rm bg}$&  $T_{\rm bg}$ &  $v_{\rm L_1}$\\
 & (\mdot)& (\mdot)& (g cm$^{-3}$)& (dyne~cm$^{-2}$)& (K)& (cm s$^{-1}$)\\
    \hline
         sim-0 &  2.3 $\times\ 10^{-4}$&  9.7$\times 10^{-2}$&  2.6$\times 10^{-22}$&  8.9$\times 10^{-11}$&  1.0$\times 10^{5}$&  7.7$\times 10^{4}$\\
         sim-mdot-1&  1.4 $\times 10^{-8}$&  1.4 $\times 10^{-8}$ &  2.6$\times 10^{-22}$&  8.9$\times 10^{-11}$&  1.0$\times 10^{5}$&  7.7$\times 10^{4}$\\
         sim-mdot-2&  1.1 $\times 10^{-5}$&  6.4$\times 10^{-5}$&  2.6$\times 10^{-22}$&  8.9$\times 10^{-11}$&  1.0$\times 10^{5}$&  7.7$\times 10^{4}$\\
         sim-mdot-3&  3.7 $\times 10^{-3}$&  9.7$\times 10^{-2}$&  2.6$\times 10^{-22}$&  8.9$\times 10^{-11}$&  1.0$\times 10^{5}$&  7.7$\times 10^{4}$\\
         sim-vel-1& 2.3 $\times 10^{-4}$& 9.7$\times 10^{-2}$& 2.6$\times 10^{-22}$& 8.9$\times 10^{-11}$& 1.0$\times 10^{5}$&5.7$\times 10^{5}$\\
         sim-vel-2& 2.3 $\times 10^{-4}$& 9.7$\times 10^{-2}$& 2.6$\times 10^{-22}$& 8.9$\times 10^{-11}$& 1.0$\times 10^{5}$&6.2$\times 10^{5}$\\
         sim-bgT-1&  2.3 $\times 10^{-4}$&  9.7$\times 10^{-2}$&  2.6$\times 10^{-22}$&  8.9$\times 10^{-12}$&  1.0$\times 10^{4}$&  7.7$\times 10^{4}$\\
         sim-bgT-2&  2.3 $\times 10^{-4}$&  9.7$\times 10^{-2}$&  2.6$\times 10^{-22}$&  8.9$\times 10^{-10}$&  1.0$\times 10^{6}$&  7.7$\times 10^{4}$\\
    \hline
    \end{tabular}
    \label{tab:sims}
\end{table*}

We start the hydrodynamic simulation at t$_{\rm{ini}}$~=~29\,500~yr after the start of the mass transfer in the {\sc mesa} simulation and let the hydrodynamic simulation run for 21 yr, with a final mass transfer rate of $\dot{M}_{\rm fin}=9.7\times10^{-2}$~\mdot. We also performed eight different simulations, altering different physical and numerical parameters (see Table~\ref{tab:sims}). We will justify the need for these additional simulations in Section~\ref{sec:physicalandnumericalparam}.

The size of the computational domain depends on the distance between the first Lagrange point and the accreting star, initially at -d$_{\rm L_1}$~=~83.1~\rsun\ (5.79$\times 10^{12}$~cm). Hence, the computational box has dimensions d$_{\rm L_1}\leq x\leq1.25$d$_{\rm L_1}$, 
 -1.5d$_{\rm L_1}\leq y\leq1.5$d$_{\rm L_1}$ and -0.5d$_{\rm L_1}\leq~z~\leq~0.5$d$_{\rm L_1}$ with outflow boundary conditions at each of the six faces (except for the location of the nozzle that is technically \referee{an inflow} boundary). The grid was modified to have better resolution surrounding the companion star and around the nozzle, avoiding numerical problems. We employ $(72, 96, 32)$ cells at the coarsest refinement level, with three levels of refinement (see Section~\ref{sec:convergence} for convergence tests), corresponding to a maximum resolution of  $\Delta x = \Delta y = \Delta z=$~0.65~\rsun\ (4.53$\times 10^{10}$~cm). 

\section{Results: Formation and evolution of the disk}
\label{sec:results}

In what follows, we give details of the simulation whose input parameters are explained in Section~\ref{sec:initialconditions}. We refer to this simulation as the ``Reference simulation" (sim-0 in Table~\ref{tab:sims}). Later, we test the results against changes in the input parameters.

\begin{figure}
    \centering
    \includegraphics[width=\columnwidth]{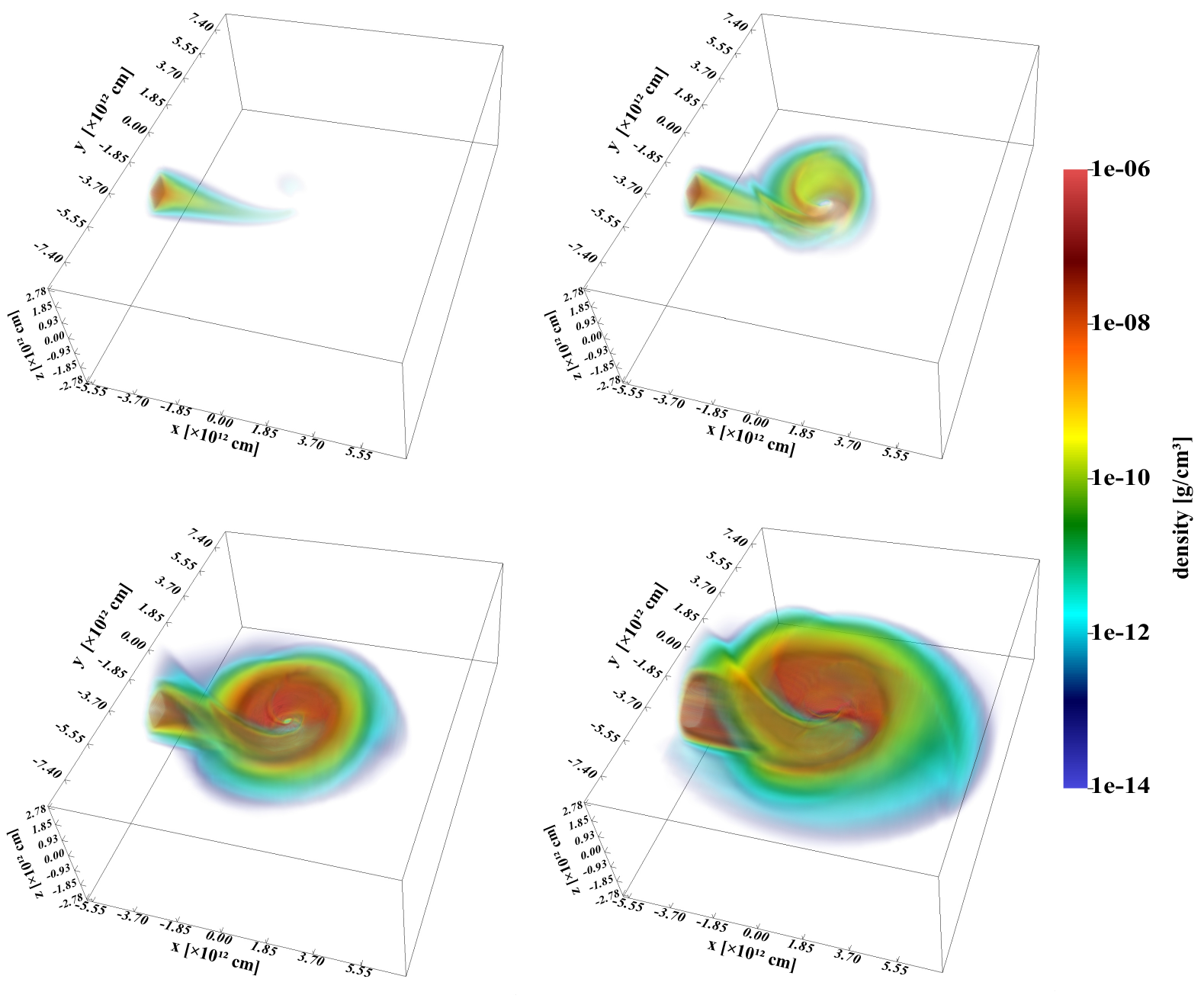}
    \caption{\referee{Volumetric density rendering showing the accretion disk at four times (t = $0.3$~yr, $1.5$~yr, $10.5$~yr, and $21$~yrs). We can appreciate the 3D structure of the accretion disk, surrounding the companion star (represented by the white area in the middle of the accretion disk).}}
    \label{fig:render}
\end{figure}

In Figure~\ref{fig:render} we show a volumetric density rendering that shows the accretion disk forming over 21~yrs. At the start of the simulation, the injected material with velocity in the $x$-direction (into the computational domain) effectively free falls towards the companion star under its own gravity, but due to its angular momentum, it sweeps around forming a disk. The injected material moves towards the companion, taking $\approx$~0.12~yr to reach the internal boundary, and is deflected around the central potential toward the lower density medium. The interaction between the deflected material and the injection stream creates the accretion disk's base structure, seen after 0.3~yr (upper left panel of Figure~\ref{fig:render}). One year later, the accumulated deflected material forms two high density spiral arms on the orbital plane (see upper right panel). 

After 10.5~yr the shock between the spiral arms has created a high density structure with a disk-like shape, surrounding the companion star (see bottom left panel). The material in the disk is orbiting around the companion star, forming bow shocks with the injected material, showing a higher density on the left side of the computational domain than on the right side, consistent with the results of \citet{makita2000two}\footnote{Since much of this work is based on the work of \citet{makita2000two} we will continue the comparison throughout this paper and bring it to bear in a discussion in Section~\ref{sec:makita}.}. By the end of the simulation at 21~yr (see bottom right panel of Figure~\ref{fig:render}) the accretion disk  has maintained its approximate structure for $~10.5$~yr, albeit while growing somewhat in radius and scale height, due to the accumulation of mass.

In the next section, we quantify what we have just described qualitatively.

\subsection{Accretion disk parameters}
\label{sec:isadisk}

\begin{figure*}
    \centering
    \makebox[\textwidth]{\includegraphics[width=\textwidth]{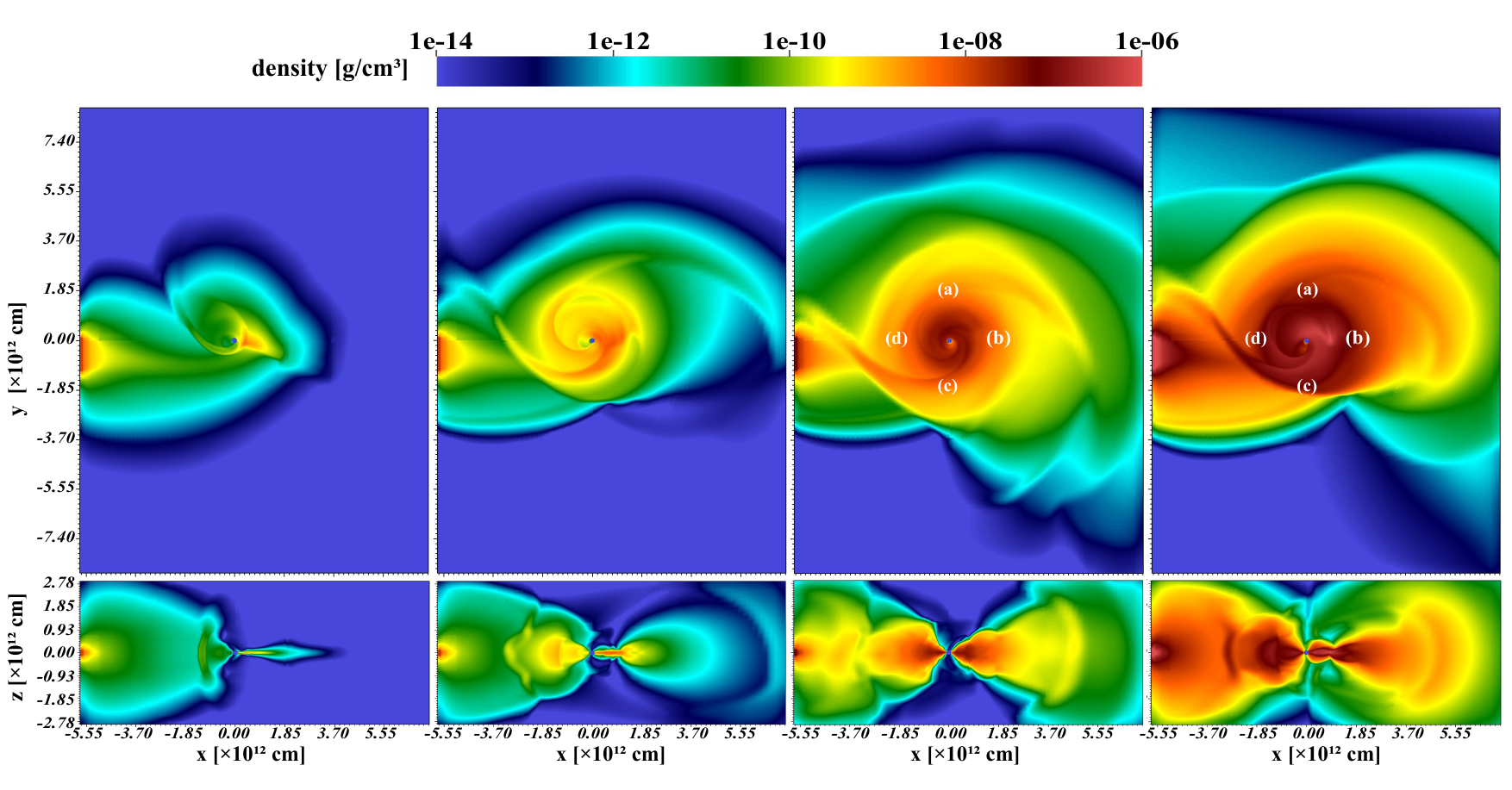}}
    \caption{\referee{Density slices of the accretion disk on the orbital plane (top row) and perpendicular plane (bottom row) of sim-0 at times t = 0.3~yr, 1.5~yr, 10.5~yr, and 20.9~yr (from left to right). The letters in the third and fourth upper panels represent the positions used to calculate the density profiles (see text).}}
   \label{fig:densitycut}
\end{figure*}
 
In Figure~\ref{fig:densitycut}, we present orbital and perpendicular density slices of the accretion disk at the same four times shown in Figure~\ref{fig:render}. Along the orbital plane we see the formation of spiral arms, the injected mass leaves the nozzle and is deflected towards the companion star, due to its gravitational pull, Coriolis force, and centrifugal force. With an increase in the rate of mass transfer, a high density structure can be seen in the last panel of the top row of Figure~\ref{fig:densitycut}. 

The edge-on panels in Figure~\ref{fig:densitycut}, bottom row, show the formation of the aforementioned bow shocks, due to the interaction between the injection stream and the accretion disk. Also, the edge-on view shows the presence of small, low density outflows along the polar axes, and may suggest the formation of hydrodynamically collimated jets in the future; however, our simulation ends too early to follow their development. These outflows are not visible in the rendered plots as a result of the choice of the colour bar limits. 

\begin{figure}
    \centering
    \includegraphics[width=0.9\columnwidth]{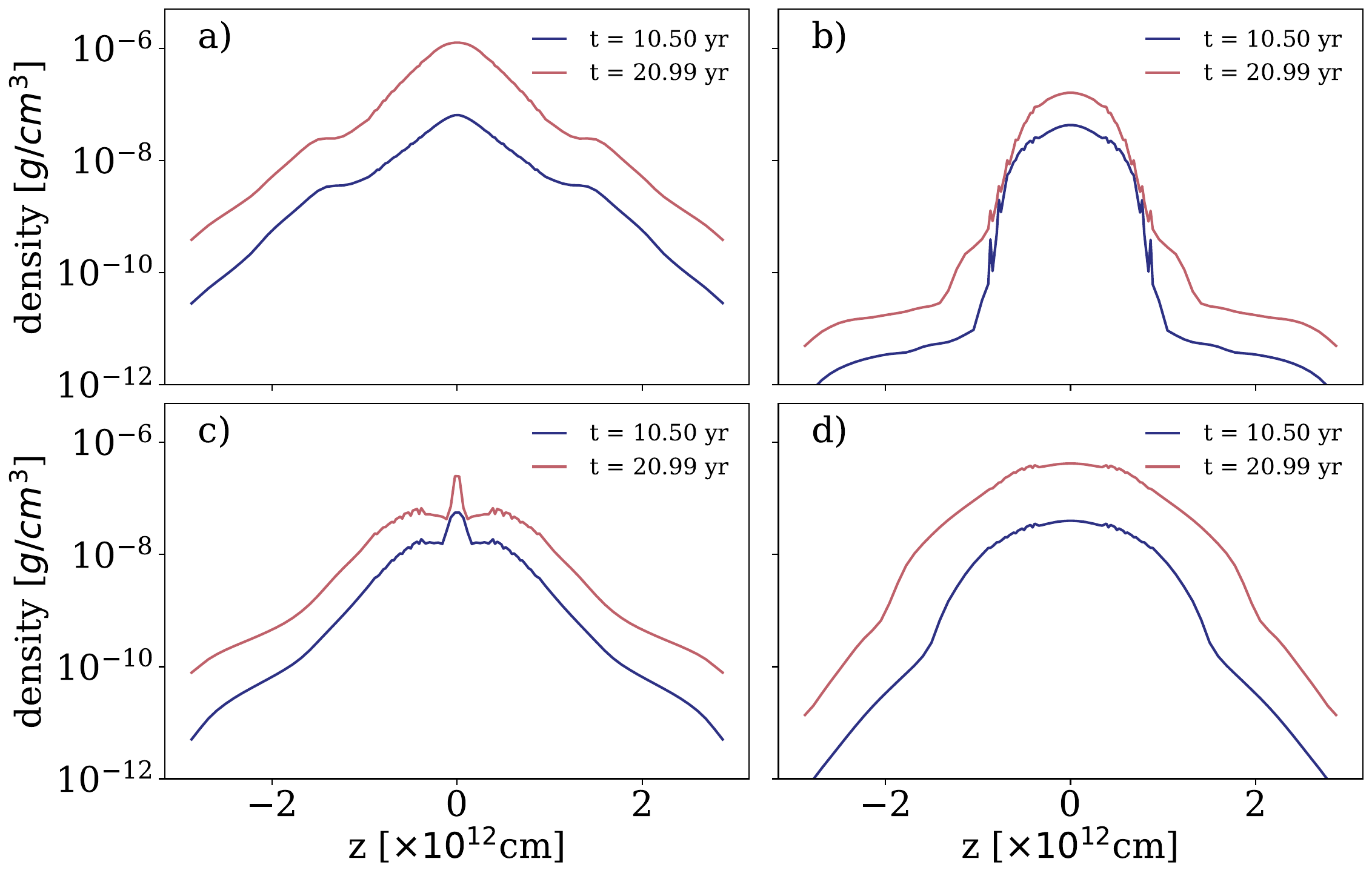}
    \caption{\referee{Density profile of the accretion disk versus height at a radius of 26.7~\rsun\ (1.86$\times10^{12}$~cm). Two times are shown: t~=~10.5~yr (blue line) and 21~yr (red line). The readings were taken at four symmetric points around the companion indicated in Figure~\ref{fig:densitycut} star along the x- and y-axis (panel (a):~+y-axis, (b):~+x-axis, (c):~-y-axis, (d):~-x-axis).}}
   \label{fig:diskheight}
\end{figure}

In Figure~\ref{fig:diskheight} we plot density profiles along the z-axis of the accretion disk, measured at 26.5~\rsun\ (1.85$\times10^{12}$~cm) from the companion star along the positive and negative side of the x- and y-axes, at t~=~10.5~yr (blue line), and t~=~21~yr (red line). \referee{The upper panels of Figure~\ref{fig:diskheight} show the density profiles measured on the positive y-axis and x-axis, panels (a) and (b), respectively. Panels (c) and (d) show the density profiles measured on the negative sides of the y-axis and x-axis, respectively (these labels are also marked in Figure~\ref{fig:densitycut})}.

Panel (d) shows the material to be more extended in the z-direction because the incoming material from the nozzle is constantly interacting with the disk material. Meanwhile, panel (b) presents a clear description of the disk thickness, with a sharply decreasing density above and below the midplane, just as we can see in the density maps (Figure~\ref{fig:densitycut}). At 10.5~yr the disk has an average thickness of 15.6~\rsun\ (1.09$\times10^{12}$~cm). By the end of the simulation (21~yr) the disk has reached an average thickness of 20.4~\rsun\ ($1.42\times10^{12}$~cm). We also measure the scale height of the disk at 10.5~yr to be  H~=~6.7~\rsun\ (0.47$\times10^{12}$~cm), while at the end of the simulation (21~yr) the disk scale height has slightly decreased to 4.9~\rsun\ (0.34$\times10^{12}$~cm).

Panels (b) and (d) in Figure~\ref{fig:diskheight} show that the thickness of the disk perpendicular to the line that joins the two stars is thicker and less defined. Panel (d), includes the interaction of the deflected material that has left the nozzle and encounters the accretion disk, just as we can appreciate in the density maps of the orbital plane. The shape of the vertical density profiles at each of the four locations does not change significantly over 10 years; only the density increases over time due to the constant injection of material.

\begin{figure}
    \centering
    \includegraphics[width=\columnwidth]{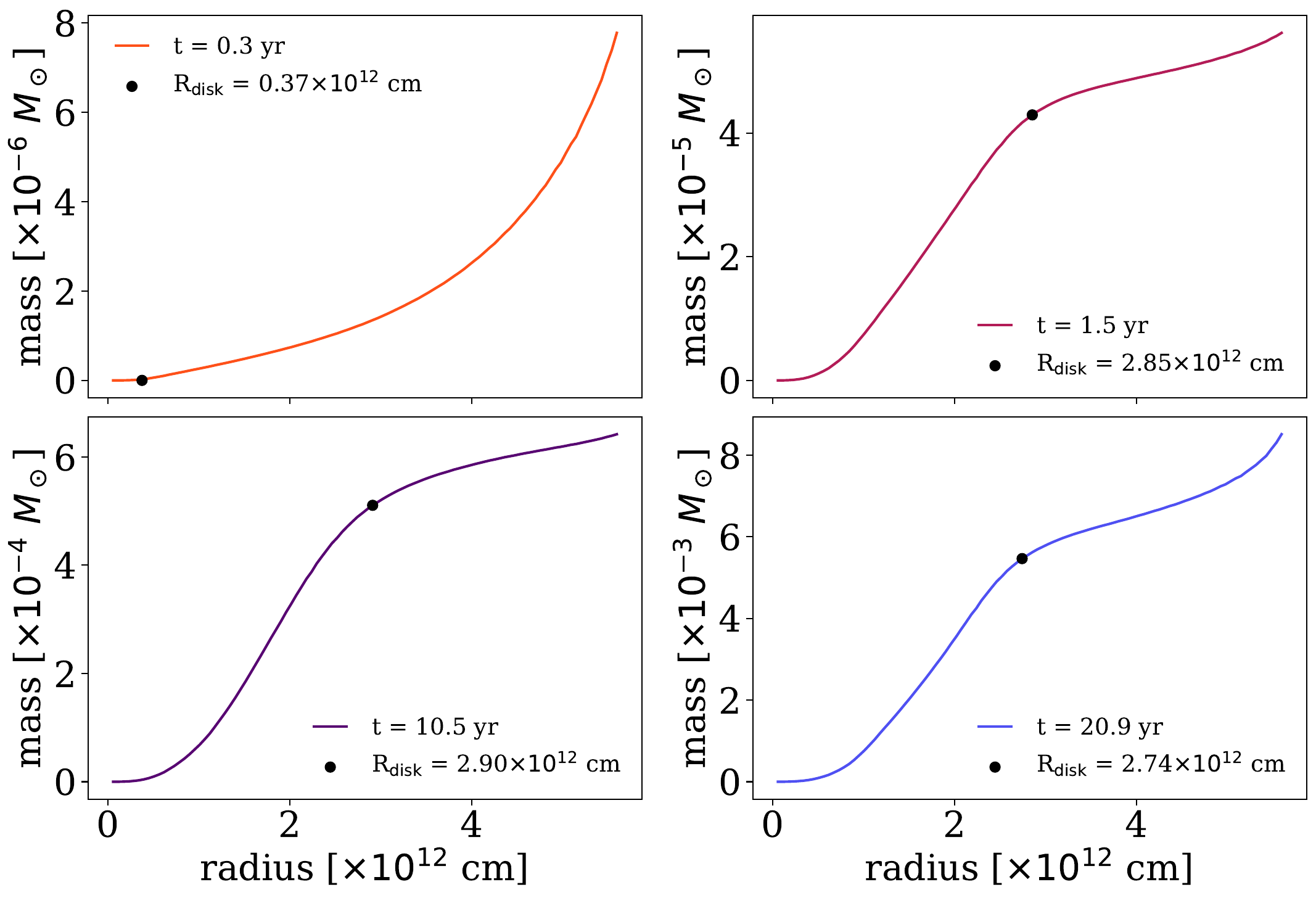}
    \caption{Cumulative mass as a function calculated from the companion star, at different times: t~=~0.3~yr (top left panel, orange line), t~=~1.5~yr (top right panel, maroon line), t~=~10.5~yr (bottom left panel, purple line), t~=~21~yr (bottom right panel, blue line). The black symbol in each panel indicates the adopted radius of the disk.}
    \label{fig:diskmass}
\end{figure}

We measured the mass inside one hundred radial points (Figure~\ref{fig:diskmass}), measured from the inner boundary (encircling the companion star) to the edge of the computational domain. We repeat this calculation for the same  times as in Figure~\ref{fig:densitycut}. At $t=0.3$~yr (orange line panel) the injected material has just reached the internal boundary, but the disk has not formed yet. Once the disk has formed, at $1.5$~yr (maroon line panel in Figure~\ref{fig:diskmass}), the enclosed mass increases with radius until it reaches the edge of the disk. At this point, we see a flattening of the slope of the enclosed mass. The nozzle is located at a radius of 74.4~\rsun\ ($5.18\times10^{12}$~cm) from the inner boundary, and it manifests itself in the steepening of the gradient at the largest radius, particularly evident at the last time (bottom right panel in Figure~\ref{fig:diskmass}). 
We determine the radius of the disk by locating the first inflection point (black filled circle in all panels).  \referee{The cumulative mass profile and hence the inferred mass and radius of the disk, depend somewhat on assumed parameters such as injection velocity. We will explore these dependencies in Sections~\ref{sec:masstransfer} and \ref{sec:velonozzle}.} In Table~\ref{tab:results} we summarise the disk characteristics discussed so far.

\begin{table}
    \centering
    \begin{tabular}{ccccc}
    \hline
    \hline
         Time (yrs) &  R$_{\rm disk}$ (\rsun) &  M$_{\rm disk}$ (\msun) &  H$_{\rm disk}$ (\rsun) & $\dot{M}_{\rm acc}$ (\msun/yr)\\
    \hline
         0.3 &  5.3 &  2.3 $\times\ 10^{-6}$ & ---- & --- \\
         1.5 &  41  &  4.3 $\times\ 10^{-5}$ & ---- & ---\\
         10.5 & 41  &  5.1 $\times\ 10^{-4}$ & 6.7 & 3.6$\times10^{-4}$\\
         20.9 & 39  &  5.5 $\times\ 10^{-3}$ & 4.9 & 3.7$\times10^{-3}$ \\
    \hline
    \end{tabular}
     \caption{\referee{Disk properties at different moments in time in sim-0.}}
    \label{tab:results}
\end{table}

\begin{figure}[ht]
    \centering
    \includegraphics[width=0.9\columnwidth]{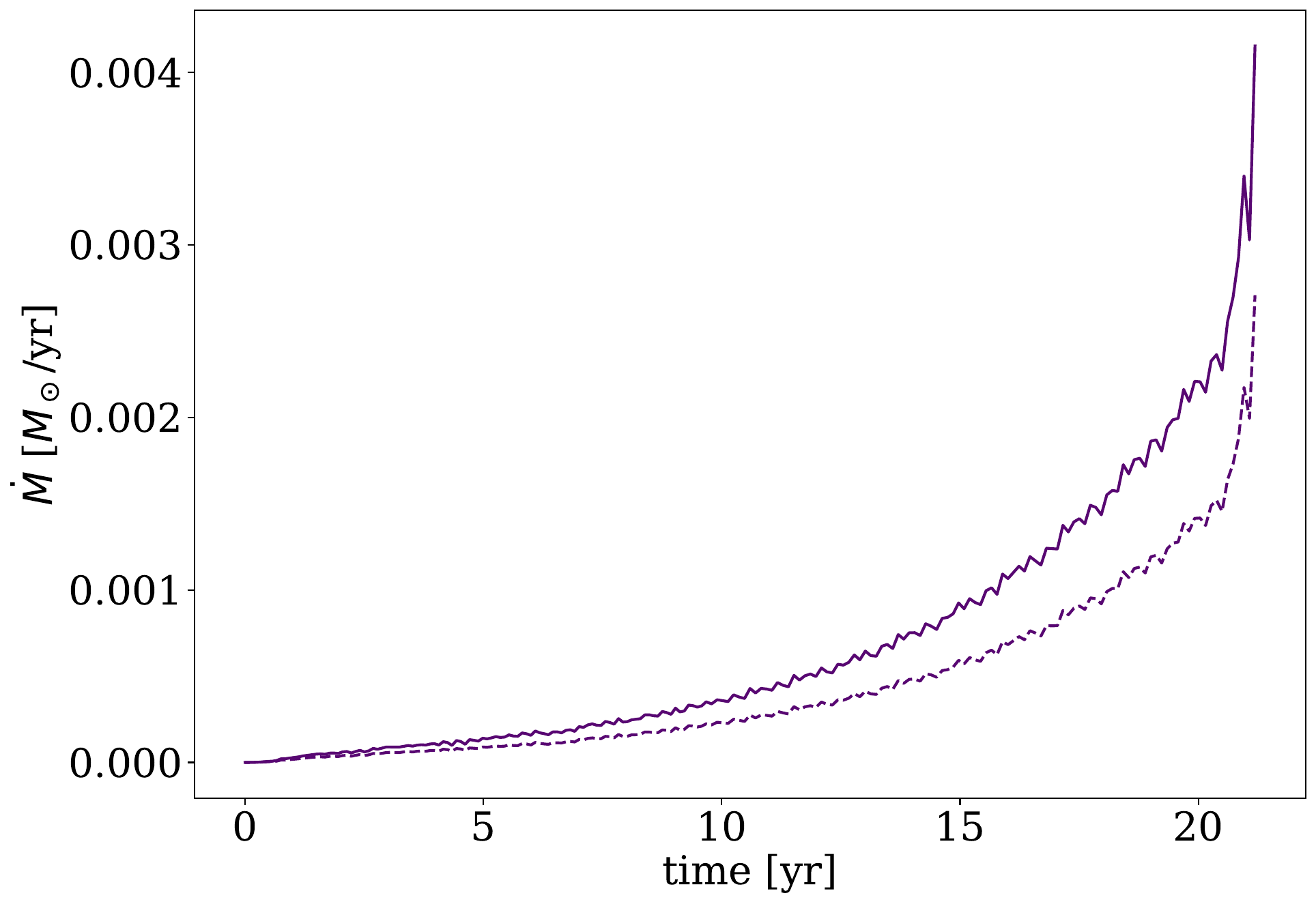}
    \caption{Mass accretion rate onto the companion (traversing the inner, inflow boundary) as a function of time. The solid line represents the mass accretion rate assuming the inner boundary is a cube with edge length 2R$_{\rm{in}}$ and the dashed line shows the mass accretion rate calculated projecting the area and velocity of the cell over the spherical inner boundary.}
    \label{fig:massaccreted}
\end{figure}

Using the inner inflow boundary, we measure the accretion rate onto the companion as a function of time (plotted in Figure~\ref{fig:massaccreted}). We identify a cubic volume, 2$R_{\rm in}$ on a side, which contains the inflow boundary sphere. The total accretion rate is measured by calculating the mass flux through the cubic boundary, by taking the projection of the velocity of each cell surrounding the boundary onto the normal direction to the boundary (solid line in Figure~\ref{fig:massaccreted}). If the projected velocity of a cell is pointing inside the fixed box, \referee{we assume that the material in that cell is  accreted and deleted from the simulation in the next timestep},  or  
\begin{equation} \dot{M} = \sum_{\rm{i=1}}^{\rm{N}}  \rho_{\rm{i}} A_{\rm{i}}\vec{v}_{\rm{i}} \cdot \hat{n},  \end{equation}
where $\rho_{\rm{i}}$ is the density of the cell, $\vec{v}_{\rm{i}}$ is the velocity vector of the cell, $A_{\rm{i}}$ represents the area of the cell face through which the material will cross in the next time step, and $\hat{n}$ is the vector perpendicular to that face directed into the cubic boundary. We also carry out the same measurement with a slightly different approach: we project the velocity vector of each cell as well as the area of the face that the gas crosses, along the radial direction to the companion -- this method results in somewhat smaller values for the accretion rate by a factor of 0.6 (dashed line in Figure~\ref{fig:massaccreted}).

\referee{The values of the accretion rate determined with the first method above are listed in Table~\ref{tab:results}. The accreted mass does not contribute significantly to the mass of the companion, which increases only by 8.4$\times10^{-3}$~\msun\ during 21 years.}
  
\referee{The accretion rate into the inner boundary in our simulations may be affected by numerical viscosity. However, the fact that the accretion rate values are approximately 100 times smaller than the injection rate at $L_1$, argues that numerical viscosity does not play a substantial role. We later also show that the accretion rate is well converged (Section~\ref{sec:convergence}), in agreement with this conclusion. It would therefore be reasonable to state that viscosity in our simulations is provided by disk turbulence and possibly the effect of spiral shocks. We can therefore compare the numerically-derived accretion rates to those theoretically predicted by \citet{Shakura1973}}:
\begin{equation}
    \dot{M}= 3 \pi \Sigma \alpha c_{\rm s} H_{\rm{disk}},
\end{equation}

\noindent where $\Sigma$ is the surface density of the disk, $\alpha$ is an efficiency factor and where we assume \linebreak $\Sigma~=~M_{\rm{disk}}/R_{\rm{disk}}^2$). For thin disks, we arbitrarily assume $\alpha=0.1$. The variable c$_s$ is the speed of sound at R$_{\rm{disk}}$ and H$_{\rm{disk}}$ is the scale height of the disk. Using the aforementioned results \referee{at the 10.5 and 21 years, we predict accretion rates of $5.0\times10^{-4}$~\mdot and $3.7\times10^{-3}$~\mdot, respectively, consistent with the accretion rate measured at the same two points in the simulation (Table~\ref{tab:results}). }

\begin{figure}[h]
    \centering
    \includegraphics[width=0.93\columnwidth]{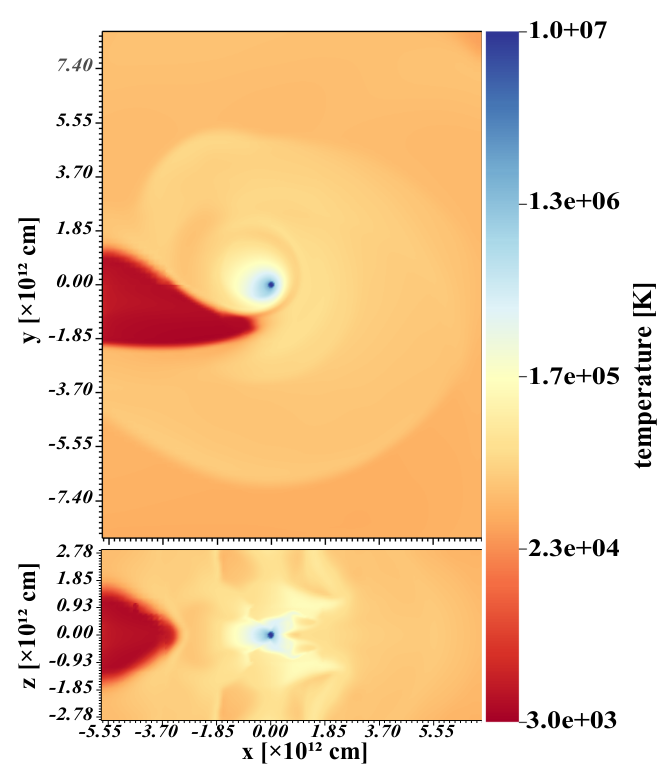}
    \caption{Temperature slices for sim-0 in the orbital (top panel) and perpendicular (bottom panel) planes of the disk at t~=~21~yr.}
    \label{fig:disktempslices}
\end{figure}

The material in the nozzle has density ($\rho_{\rm{L_1}}$) and pressure ($P_{\rm{L_1}}$) defined in terms of the mass injection rate, the volume of the nozzle, and the velocity, as explained in Section~\ref{sec:initialconditions}. Assuming an ideal gas, we show, in Figure~\ref{fig:disktempslices}, temperature slices in the orbital and perpendicular planes at the end of the simulation (21~yr). The  material inside the nozzle has an average temperature of $\sim 5\,000$~K, once the material leaves the nozzle its temperature drops to $\sim 3\,000$~K, due to the pressure difference between the nozzle and the material just outside the nozzle. Inside the inflow boundary around the accretor, the high temperature ($\sim 10^{7}$~K) is due to the imposed low density. We observe a gradient of temperature decreasing radially away from the inner boundary, we can also observe on the perpendicular plane (x-z plane) the interaction between the injected mass  and the mass that circles around the companion.

\begin{figure}
\centering
\includegraphics[width=\columnwidth]{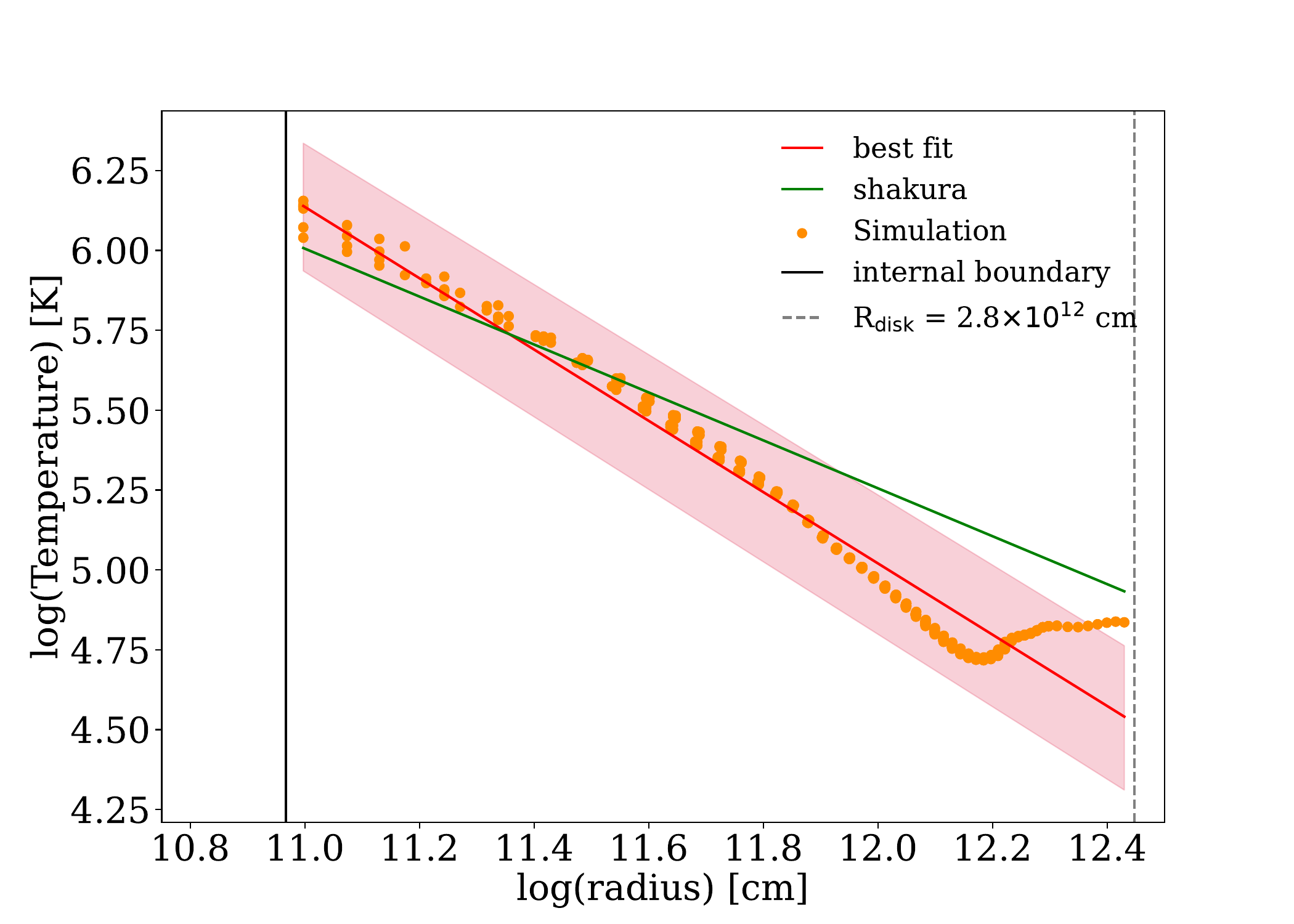}
\caption{Temperature profile for sim-0 in the mid plane along the positive x-axis at t~=~21~yr. The disk's radius is indicated with a vertical grey dashed line, while the inner boundary's radius is marked with a vertical black line. The solution for steady disks is indicated by the green curve ($T~\propto~r^{-3/4}$) and the best fit ($\alpha = 1.12\pm0.02$) is indicated by the red curve and red shaded area. The average percentage error between the temperature profile and the fit line is 15\%. }
\label{fig:disktemperature}
\end{figure}

In Figure~\ref{fig:disktemperature}, we plot a temperature profile along the positive x-axis at 21 years, with the internal boundary shown as a black vertical line and the radius of the disk as a dashed grey line. In the mid-plane along the positive x-direction, the cells closer to the inner boundary have a temperature of $1.1\times10^{6}$~K, their location corresponds with the region where the pressure gradients is the highest; the cells located at the edge of the disk have lower temperatures of $\sim$60\,000~K . 

Following the prescription of \cite{shakura1973black} for steady disks, the effective temperature profile should follow $T\propto r^{-0.75}$ (green line in Figure~\ref{fig:disktemperature}). This function fits our data with an average percentage error of 38\%. The best fit is obtained using a steeper $T\propto r^{-1.12\pm 0.02}$ (red line and shaded area in Figure~\ref{fig:disktemperature}) with an average percentage error of 15\%. \cite{Mineshige94} stated that a value of $\alpha < -0.75$ characterises X-ray binary systems in the flaring branch on the X-ray hardness-intensity diagram. 

\referee{However, our simulation does not have explicit cooling, so we may expect it to diverge from the prediction of \citet{Shakura1973}.  Instead, our adopted equation of state implies that $T \propto (\gamma-1)e$, and as the energy dissipated per unit mass in an accretion disk scales as $1/r$, it is natural for the temperature to scale similarly.}

\begin{figure}
\centering
\includegraphics[width=\columnwidth]{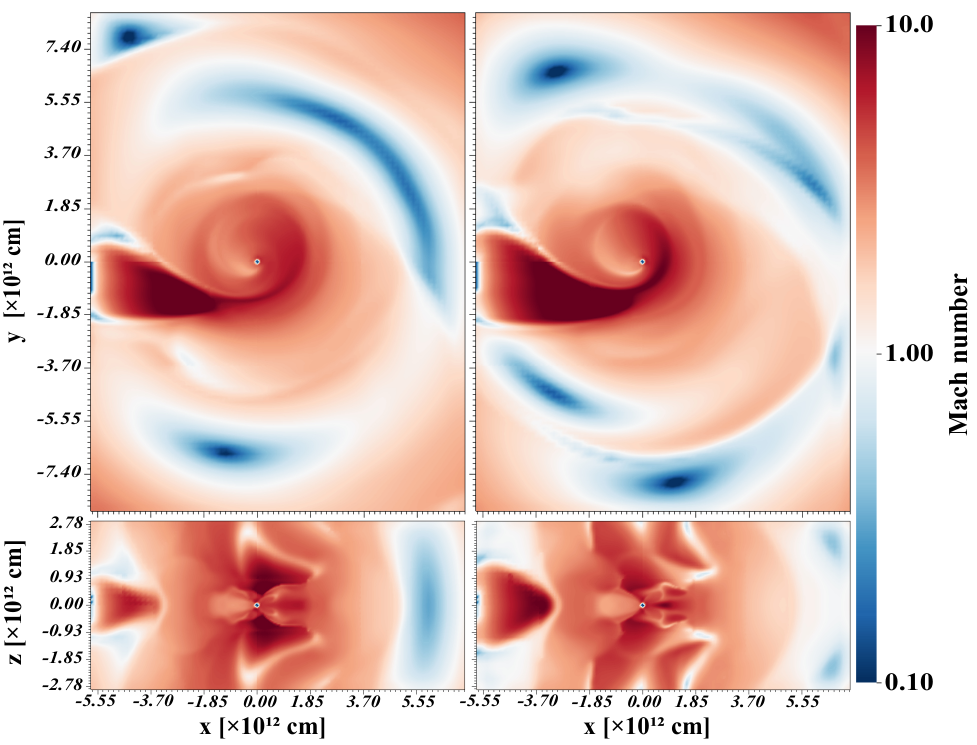}
\caption{Slices of Mach number in the orbital plane (top panels) and perpendicular plane (bottom panels) of the disk at t~=~10.5~yr (left panels) and t~=~21~yr (right panels) for sim-0. 
}
\label{fig:diskvelocity-mach}
\end{figure}

The material is injected through the nozzle with a subsonic velocity of $7.7\times10^4$~cm~s$^{-1}$ in the x direction (note that the sound speed value near the nozzle ranges between 6 and 8~km~s$^{-1}$). In the upper panels of Figure~\ref{fig:diskvelocity-mach}, we show the Mach number at 10.5~yr and at the end of the simulation (21~yr), with slices in the orbital and perpendicular planes. The nozzle is on the left of the domain, seen as a short vertical bar with subsonic velocity. Just outside the nozzle, the velocity becomes highly supersonic ($\mathcal{M}$~=~27 at 21~yr). When the injected material approaches the centre of the domain, it circles it and collides with the material that was already in orbit around the companion star, slowing down.

\begin{figure}
\centering
\includegraphics[width=0.9\columnwidth]{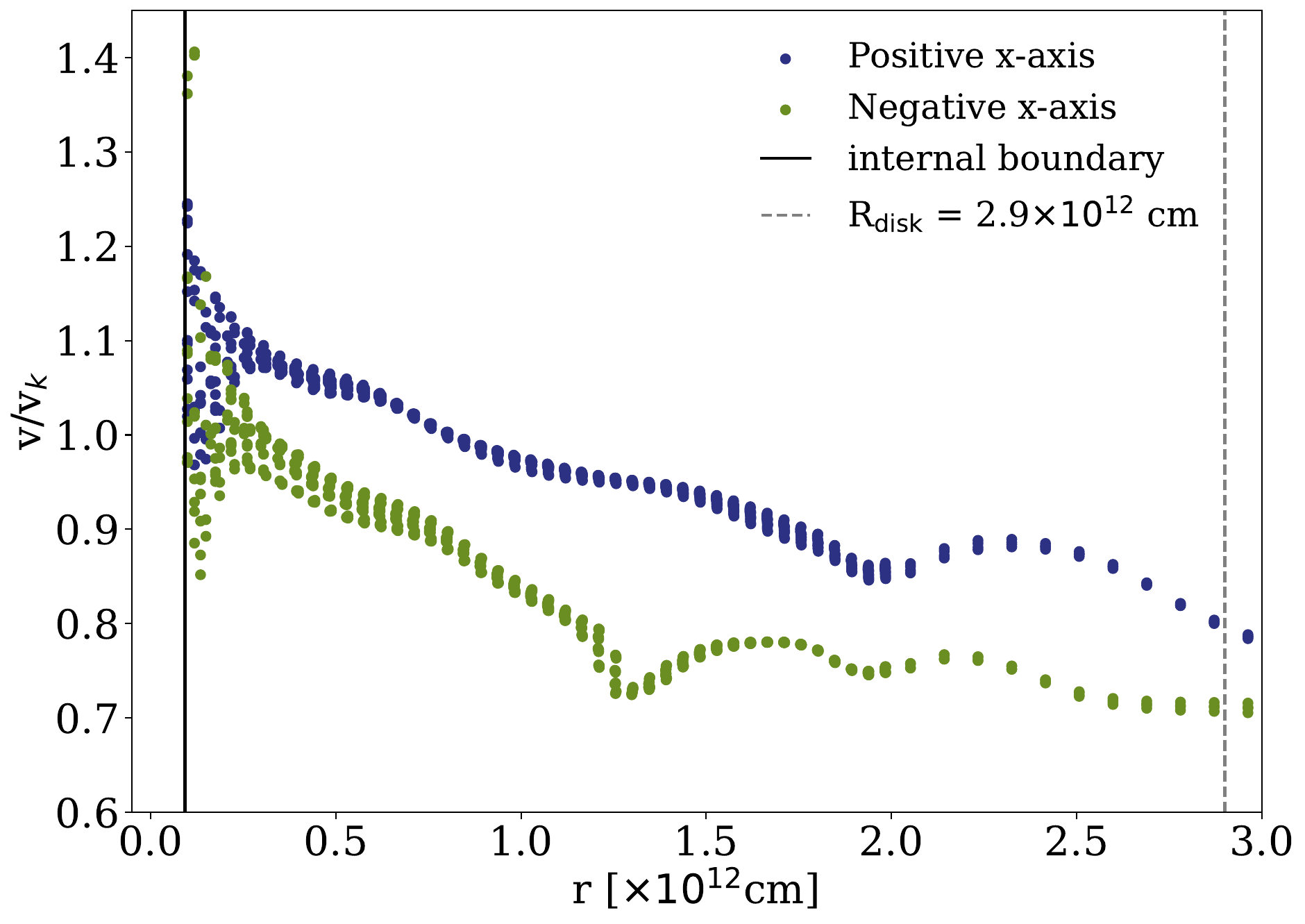}
\caption{Velocity profile in the mid-plane along the positive and negative axes at $10.5$~yr. The velocity is  normalised to the Keplerian velocity (v$_{\rm k}$). The inner boundary is marked by the black vertical solid line and the radius of the disk by the vertical grey dashed line.}
\label{fig:diskvelocity}
\end{figure}

In Figure~\ref{fig:diskvelocity}, we plot the gas velocity profile in the mid-plane along the positive and negative x-axes at t~=~10.5~yr, normalised to the Keplerian velocity (v$_{\rm k}$~=~$\sqrt{GM/r}$). \referee{We note that normalising velocities to the Keplerian values is appropriate because gas velocity vectors are effectively entirely in the azimuthal direction, except for a few cells near the centre that have a 20\% radial component.}  The inner boundary radius, R$_{\rm{in}}$, is marked with a black line, while a grey line indicates the radius of the disk, R$_{\rm{disk}}$. The cell velocities around the inner boundary are Keplerian on average, though there is quite a bit of scatter in individual cells \referee{due to the gas there not moving entirely azimutally}. Cells at the edge of the disk move with velocities approximately 80\%~-~90\% of the Keplerian value. This is likely due to the pressure support of the gas in the disk. 


\begin{figure*}
    \centering
    \includegraphics[width=\columnwidth]{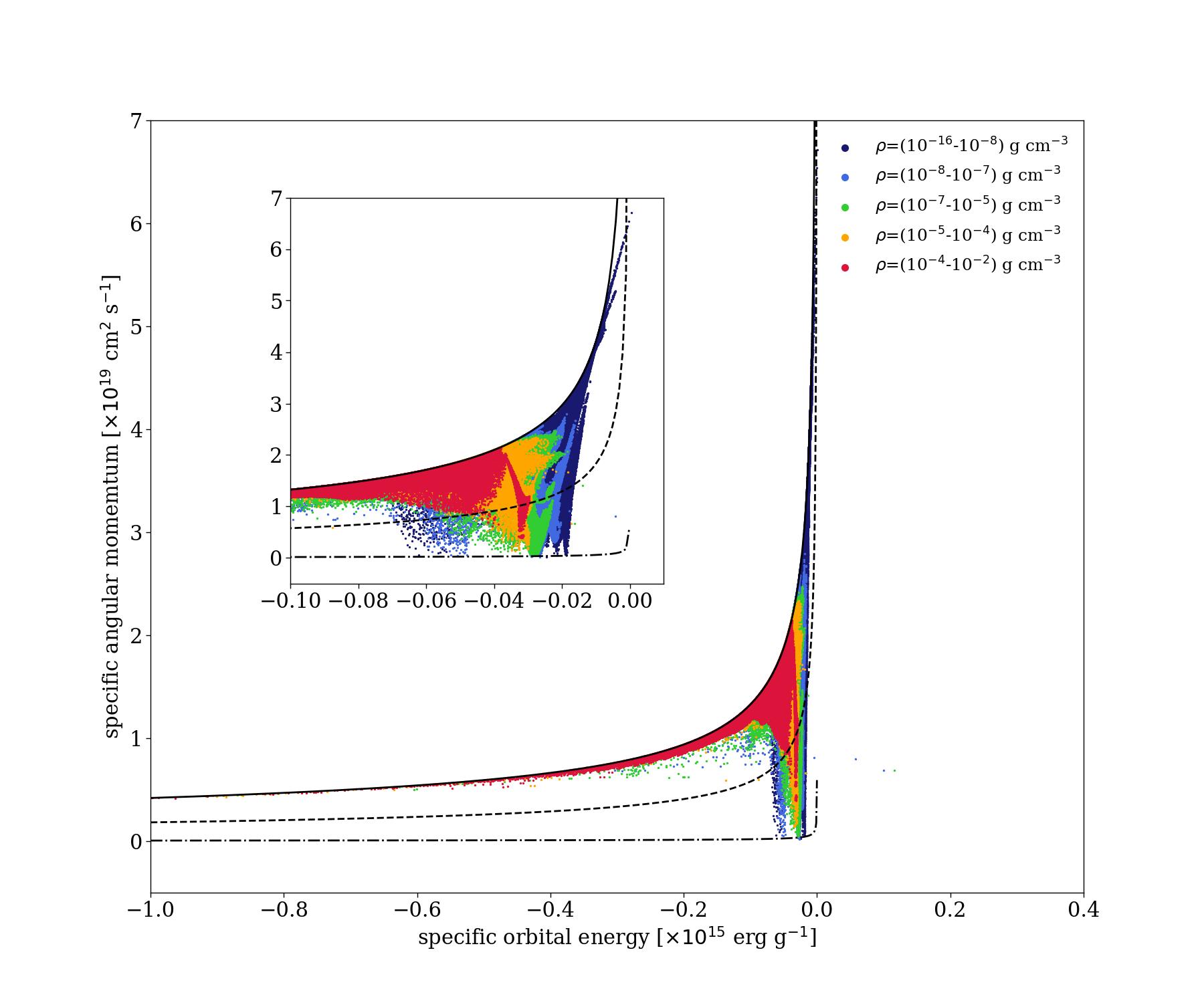}
    \caption{Specific angular momentum versus specific orbital energy for every cell in the computational domain once the disk has formed (sim-0). The integration time corresponds to t~=~10.5 yrs. The density colour table is the same as in Figure~\ref{fig:densitycut}. The black line is the solution for a circular orbit. The black dashed line and the black dash-dotted line represent orbits with higher eccentricity (e~=~0.9 and 0.9999, respectively). The time evolution of this plot is available at \url{https://drive.google.com/file/d/10Kh7uXEJaRr8zY5GNmyQr24Vy9wC--G-/view?usp=sharing}.}
    \label{fig:jvse}
\end{figure*}

We further study the disk by examining the locus of the gas cells in the specific energy vs. specific angular momentum plane \citep[][Figure~\ref{fig:jvse}\footnote{The time evolution of the specific angular momentum versus energy plot is available at \url{https://drive.google.com/file/d/10Kh7uXEJaRr8zY5GNmyQr24Vy9wC--G-/view?usp=sharing}}]{2021Hayashi}. We calculate the specific orbital energy as 

\begin{equation}
    e_{\rm{orb}} = {\frac{1}{2}} v^2 - {\frac{GM}{r}},
\end{equation} 

\noindent where $v$ is the magnitude of the velocity in the centre of the cell, $r$ is the position of each cell with respect to the companion star, and $M$ is the mass of the companion star. The magnitude of the specific angular momentum for each cell is:
\begin{equation}
   j_{\rm{orb}} = r \cdot v.
\end{equation}

Each cell is colour-coded by density (with the same colour bar as in Figure~\ref{fig:densitycut}), where the cells that compose the high density accretion disk are marked in red, and dark blue points represent the low density medium cells. The black line in Figure~\ref{fig:jvse} represents the analytical solution of a particle moving on a circular Keplerian orbit, while the black dashed line and the back dash-dotted line indicate orbits with eccentricity $0.9$ and $\sim 1$, respectively. 

The majority of the high-density cells have negative orbital energy, which means that the material is bound to the companion star. On the other hand, the insert plot, showing a zoom-in near the origin, shows that some of the material has positive orbital energy; hence it is unbound. Some of the gas dropped by the nozzle onto the companion star is not initially bound to the system but slows down upon colliding with gas ahead of it. The cells in the high-density region are dispersed between the black line and the grey dashed line, which means that they move in closed orbits with eccentricities less than 0.9. \referee{The gas behaviour in this diagram is consistent with disk formation.}

\referee{We finally determine whether the viscosity that enacts such large accretion rates is consistent with reasonable magnetic fields (noting that our simulations do not have magnetic fields and that in the simulations the viscosity is enacted by turbulent gas motion and possibly shocks). Using $10^{15}$~G at the surface of the neutron star \citep{Rea2025}, one would find that the field at 1.3~\rsun\ is of the order of a Gauss, following magnetic flux conservation for a dipole ($B\sim r^{-3}$). Such low magnetic flux is unlikely to be responsible for the accretion rates that we measure (or that we infer using the formalism of \citet{Shakura1973}). In fact using the expression of  \citet{Wardle2007}:}

\begin{equation}
    \dot{M} \simeq 6\times10^{-12}\left(\frac{B}{1G} \right)^{2} \left(\frac{r_{\rm in}}{ 1.3R_\odot}\right)^{5/2} \left(\frac{M_{\rm NS}}{ 1.4M_\odot}\right)^{-1/2} {\rm M}_\odot {\rm yr}^{-1},
\end{equation}

\noindent \referee{where $\dot{M}$ is the mass accretion rate, $B$ is the magnetic field, $r_{\rm in}$ is the accretion radius and $M_{\rm NS}$ is the mass of the accretor, the deduced accretion rate would be very low indeed. To obtain instead a mass accretion rate of the order of $4\times 10^{-3}$~\msun~yr$^{-1}$, as is measured at the end of the simulation, would necessitate a field at the inner disk rim of 28\,000~G. Such field could possibly be obtained by  magneto-rotational instability amplification of a Gauss level field \citep{Balbus1991}, over as little as $\sim$2 orbital periods of the gas at the inner rim (the amplification grows as $\exp(3/4 \Omega t)$, where $\Omega$ is the orbital frequency and $t$ is time). This would argue for a need to carry out these simulations in a magneto-hydrodynamics regime \citep[e.g.][]{2020ApJ...904...90P} }.\\\\

\section{Sensitivity of results to the choice of some physical and numerical parameters}
\label{sec:physicalandnumericalparam}

\subsection{Mass injection rate}
\label{sec:masstransfer}

The full evolution of the mass transfer rate from the moment of Roche lobe overflow to the moment of CE as seen in the 1D, {\sc mesa} simulation lasts 30\,000~yr (Figure~\ref{fig:mesamdot}) and goes from $1.6\times 10^{-9}$~\msun~yr$^{-1}$ to $9.7\times 10^{-2}$~\msun~yr$^{-1}$ (Table~\ref{tab:MESA parameters}). This entire period cannot be modelled in 3D which by necessity can only simulate a much shorter time (21 years for us). The choice was therefore made to model a period of time towards the end of the 30\,000 years modelled in 1D between 29\,500 and 29\,521 years, when the mass injection rate goes between $2.3\times 10^{-4}$ and $9.7\times10^{-2}$~\mdot\ (see sim-0 in Table~\ref{tab:mdotdiff}).  

Using sim-0 as reference, we computed three additional simulations with different initial mass transfer rates (called ``sim-mdot-$\#$'' in Table~\ref{tab:sims} and Table~\ref{tab:mdotdiff}) and hence different start times in the context of the {\sc mesa} simulation. Therefore, each 3D simulation samples a different part of the $\dot{M}$ vs. time curve in Figure~\ref{fig:mesamdot}.
In Table~\ref{tab:mdotdiff} we present a summary of the initial and final parameters of these simulations, noting that, besides the mass transfer rate, all other parameters are the same as the Reference simulation sim-0 -- see Section~\ref{sec:initialconditions}).

\begin{figure*}
    \centering
    \includegraphics[width=\columnwidth]{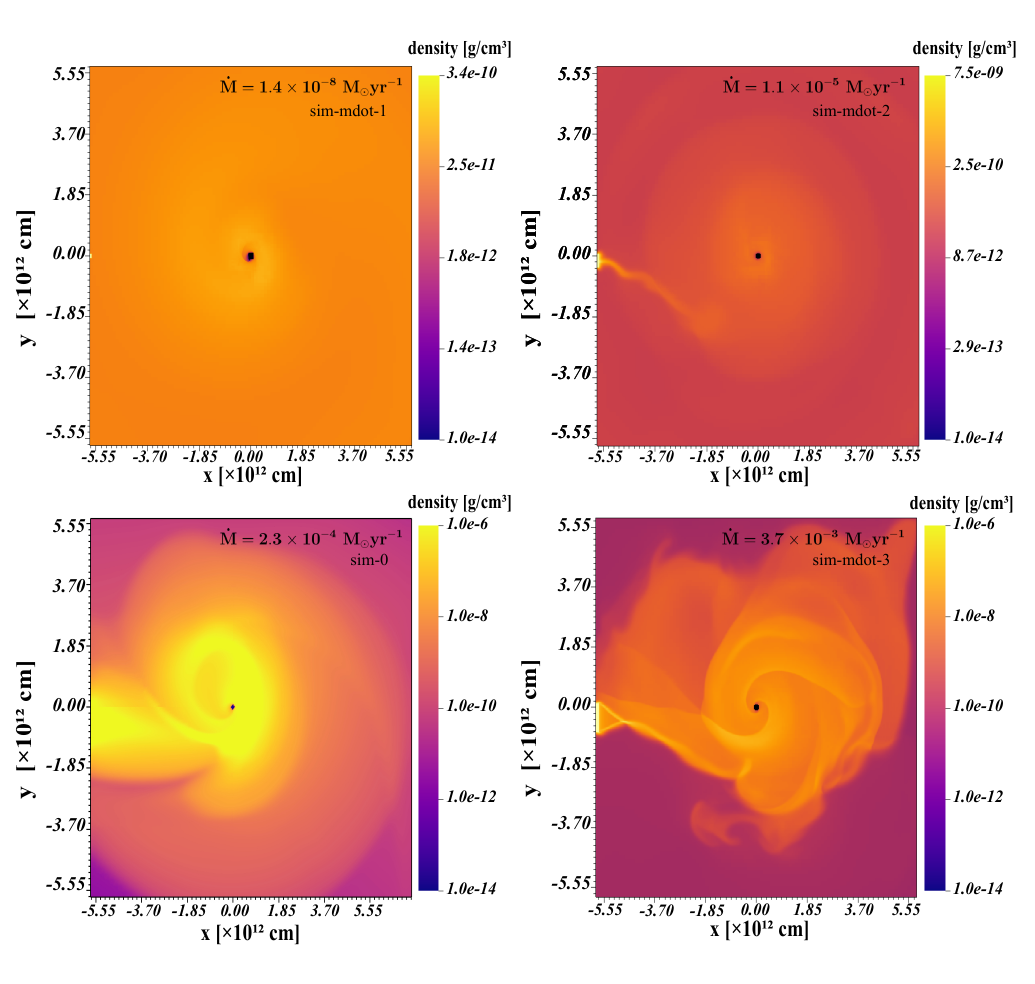}
    \caption{Densities slices on the orbital plane at the end of the 3D simulations. The models that are shown are: sim-0 (lower left panel), sim-mdot-1 (upper left panel), sim-mdot-2 (upper right panel) and sim-mdot-3 (lower right panel). Note how the density colour bar may have different maximum limits.}
    \label{fig:mdotdiff}
\end{figure*}

In Figure~\ref{fig:mdotdiff}, we present density slices in the orbital plane of these 3D simulations at their final time step (see the fifth column in Table~\ref{tab:mdotdiff}). In the upper left panel, we show the simulation sim-mdot-1, with the smallest initial mass transfer rate of $1.4\times10^{-8}$~\mdot, running for the first 128 years of the 30\,000~yr {\sc mesa} simulation. As expected, only a small, low mass disk forms at such low mass transfer rate, even if the run time is relatively long.

In sim-mdot-2 with initial mass transfer rate of $1.1~\times10^{-5}$~\mdot\ (top right panel), starting towards the end of the 30\,000-yr long {\sc mesa} simulation (only 120~yr before sim-0) and running for 29~yr, a disk forms with a radius of just under 50~\rsun\ and a mass that is not that dissimilar to that of sim-mdot-1, despite the mass injection rate being larger by three orders of magnitude (the run time was four times smaller). 

In the lower left panel of Figure~\ref{fig:mdotdiff}, we present sim-0, our reference simulation. The initial mass accretion rate is only 20 times higher at the start than the previous simulation, sim-0, but it increases far more steeply than for sim-mdot-2. The disk size is of the same order as the one in sim-mdot-2, but the disk mass is $\sim$100 times larger. 

Finally, sim-mdot-3 is very similar to sim-0 but starts 8 years later with an initial mass transfer rate that is again, $\sim$20 times larger. It runs only 13 years to the same end point as sim-0 (see the lower right panel, Figure~\ref{fig:mdotdiff}). The disk reaches a similar radius ($\sim 43$~\rsun) and a mass that is 1.6 times larger than the mass of the accretion disk of sim-0, despite the fact that the simulation started 8 years later and therefore ran for only 13 years compared to 21 of the reference simulation and injected slightly less mass. This is due to the fact that disk growth is not only dependent on the mass transfer rate and length of simulation, but also on the specific geometry of the flow, which dictates how much mass accretes through the inner boundary as well as the shape of the disk at the time it is measured, whereby the ``edge" of the disk as defined by our criteria (Section~\ref{sec:isadisk}) can vary slightly.

With hindsight, these tests could be performed more systematically so as to gain a better idea of whether the disk parameters as stated in Table~\ref{tab:mdotdiff} are close to what we might expect to be the disk just before CE for a systems such as ours. With these tests as they are we can only state that the disk's mass and radius are likely reasonable within a factor of $\lesssim 2$ for the mass, and 15\% for the radius. Given other sources of uncertainty this is a reasonable and sufficient statement for now.

\begin{table*}
    \centering
    \begin{tabular}{lccccccccccc}
    \hline
    \hline
         Model& $\dot{M}_{\rm L1, start}$ &  {\sc mesa} & $\dot{M}_{\rm L1, stop}$ & Length & Injected & Donor  & Accretor  & Orbital  & R$_{\rm{L1}}$ & M$_{\rm disk}$ & R$_{\rm disk}$ \\
         & &  start time &  & of sim. & mass &  mass &  mass &  separation & & & \\
 & (M$_\odot$ yr$^{-1}$) & (yr) & (M$_\odot$ yr$^{-1}$) & (yr) &  (M$_\odot$) & (M$_\odot$) &(M$_\odot$) & (\rsun) & (\rsun) & (\msun) & (\rsun) \\
       \hline  
        sim-mdot-1 & 1.4 $\times\ 10^{-8}$  & 0       & 1.4 $\times\ 10^{-8}$ & 128     &1.8 $\times\ 10^{-6}$& 6.98 & 1.40 & 270 & 84 & 6.3$\times10^{-9}$ &  10 \\

            sim-mdot-2 & 1.1 $\times\ 10^{-5}$  & 29\,380 & 6.4 $\times\ 10^{-5}$ & 29&1.1 $\times\ 10^{-3}$& 6.98 & 1.40 & 268 & 83 & 5.1$\times10^{-5}$ &  46\\
                     sim-0& 2.3 $\times\ 10^{-4}$ & 29\,500 & 9.7 $\times\ 10^{-2}$ & 21&8.1 $\times\ 10^{-2}$& 6.97 & 1.41 & 266 & 83 & 5.5$\times10^{-3}$ &  39 \\
sim-mdot-3 & 3.7 $\times\ 10^{-3}$  & 29\,508 & 9.7 $\times\ 10^{-2}$ & 13&5.2 $\times\ 10^{-2}$& 6.94 & 1.44 & 258 & 81 & 8.6$\times10^{-3}$ &  43\\
         \hline
    \end{tabular}%
    \caption{Parameters of a sequence of 3D simulations aiming to assess the resilience of disk parameters ($M_{\rm disk}$ and $R_{\rm disk}$) to the choice of mass transfer rate and simulation length. The mass-transfer rate, donor mass, accretor mass, and orbital separation are selected at the {\sc mesa} start time. The disk mass and radius are measured at the end of the simulation.}
    \label{tab:mdotdiff}
\end{table*}

\subsection{Velocity through the nozzle}
\label{sec:velonozzle}

The distribution and kinematics of the gas at $L_1$ (the ``nozzle") is calculated according to the prescription of \cite{lubow1975gas}, \cite{Ritter1988}, and \cite{Jackson2017AStars}. They used Bernoulli's principle to describe the evolution of the gas moving from the donor's surface toward L$_{\rm{1}}$. The gas above the donor photosphere moves with a velocity v$_{\rm{L_1}}$~$\ll$~c$_{\rm{T}}$, where $c_{\rm{T}}$ is the isothermal sound speed, while near L$_{\rm{1}}$, the gas is assumed to reach a velocity comparable to the isothermal sound speed, v$_{\rm{L_1}}$~$\lesssim$~c$_{\rm{T}}$. After the gas passes L$_{\rm{1}}$, due to the pressure gradient, the gas free falls supersonically into the companion's Roche Lobe. 

In our simulation, we need to set an injection velocity because otherwise gas placed in the nozzle does not enter the computational domain at the prescribed rate. This initial nozzle velocity is therefore arbitrary, and we thus need to ensure that changing its value does not affect the disk parameters. 

We test the dependency of the simulation on the velocity of injection (v$_{\rm L_1}$) at the nozzle, by executing three different simulations using the same parameters as the reference simulation, but setting different nozzle velocities in the x-direction: a subsonic velocity of 7.75$\times 10 ^{4}$~cm~s$^{-1}$ (sim-0; 0.01 in code units), the isothermal sound speed at the donor's photosphere, or  5.74$\times 10 ^{5}$~cm~s$^{-1}$ (sim-vel-1), and a supersonic velocity of  6.16$\times 10 ^{5}$~cm~s$^{-1}$ (sim-vel-2). See Table ~\ref{tab:sims} for a summary of all the simulations parameters.

\begin{figure}
    \centering
    \includegraphics[width=\columnwidth]{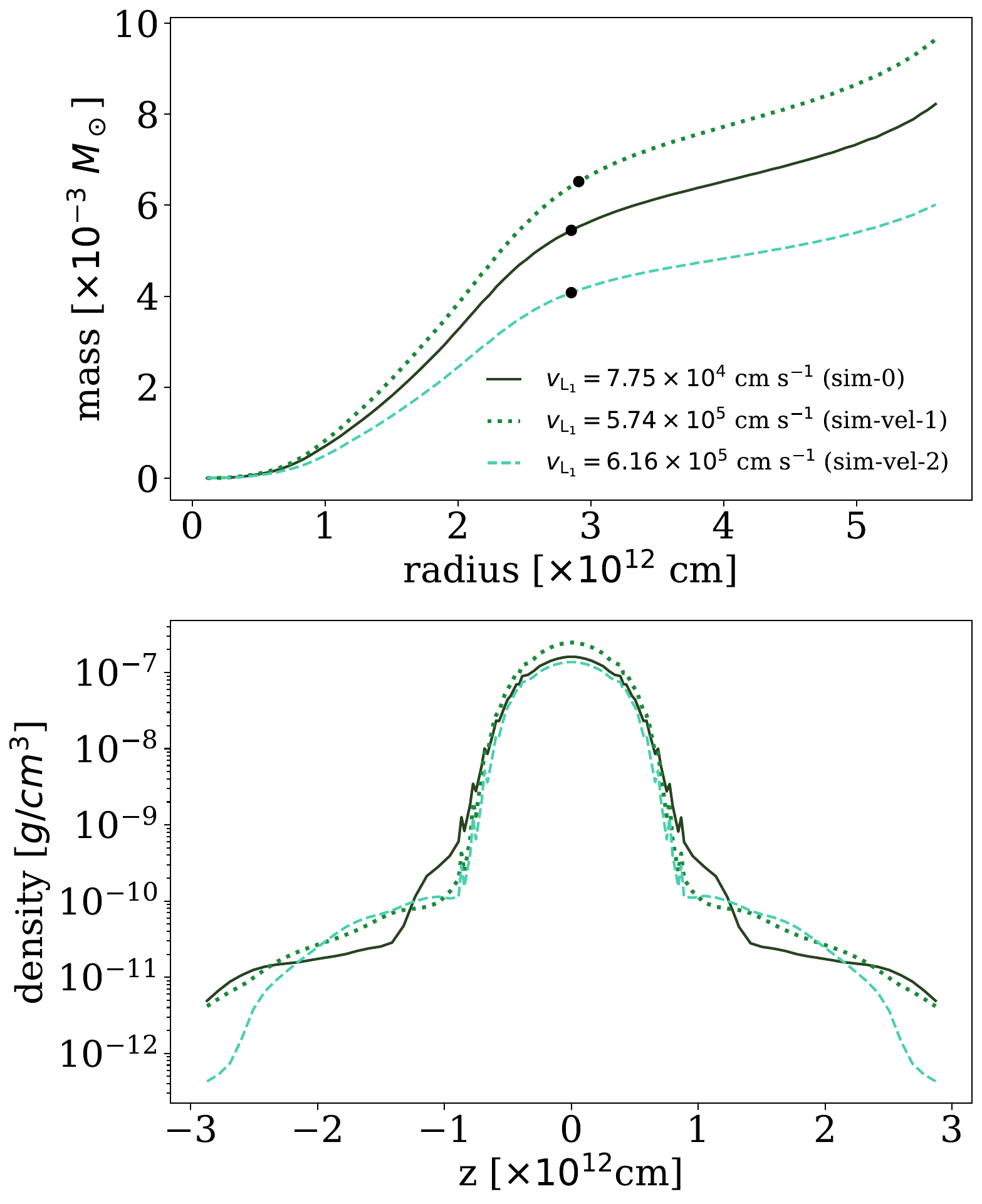}
    \caption{\referee{The cumulative mass as a function of radius (top panel) and the vertical density profile (bottom panel) for models with different injection velocities at t~=~21~yr. 
    Solid dark green line: v$_{\rm L_1}$~=~7.75$\times 10 ^{4}$~cm~s$^{-1}$ (model sim-0), dotted green line: v$_{\rm L_1}$~=~5.74$\times 10 ^{5}$~cm~s$^{-1}$ (sim-vel-1), dashed emerald line: v$_{\rm L_1}$~=~6.16$\times 10 ^{5}$~cm~s$^{-1}$ (sim-vel-2, which has the necessary velocity to leave the nozzle every time step).}}
    \label{fig:diffvelocity}
\end{figure}

The top panel of Figure~\ref{fig:diffvelocity} shows the cumulative mass as a function of radius at 21~years, similar to Figure~\ref{fig:diskmass}, the nozzle is located at a radius of 74~\rsun\ (5.2$\times10^{12}$~cm) from the inner boundary. For the reference simulation, sim-0, with the lowest injection velocity, the formation of the disk takes longer since the injected mass needs more time to leave the nozzle and to fall into the companion's Roche lobe. The accretion disk in this simulation has 11~percent more mass than simulations sim-vel-1 and sim-vel-2. We conclude that relatively small differences in the injection velocity around the isothermal velocity value c$_{\rm{T}}$, do not greatly affect the disk parameters. 

The density profile along the z-axis at 21~yr (Figure~\ref{fig:diffvelocity}, bottom panel) is similar in the three simulations at z-values close to the inner boundary, the disk scale height in the reference simulation at 21~yr (H$_{\rm{sim-0}}$~=~4.9~\rsun)
is similar to the disk scale hight for sim-vel-1 (H$_{\rm{sim-vel-1}}$~=~4.3~\rsun)
and the disk scale hight for sim-vel-2 (H$_{\rm{sim-vel-2}}$~=~5.0~\rsun).
The density at the edges of the box is an unbound low-density gas interacting with the even lower-density background; this gas does not affect the dynamics/structure of the accretion disk. The arbitrarily assumed velocity injection does not affect the final results of the simulation.

\subsection{Sensitivity of the results to background density and temperature}
\label{sec:medium}

Filling the background with low-density gas is an expedient to ensure that no cell is empty, which would cause the inability to calculate pressure gradients. The value of the background density is $\rho_{\rm bg}~=~2.60 \times\ 10^{-22}$~g~cm$^{-3}$ (1$\times10^{-16}$~code~units), the lowest viable value, below which the code does not run. As consequence of this background density choice, at the beginning of the simulation a low density rarefaction wave propagates from the discontinuity between the inner boundary and the background moving outward and out of the computational domain. The rarefaction wave leaves the computational box before the injected material reaches the inner boundary. This rarefaction wave does not affect the movement of the injected material, and the evolution of the disk formation shown in Figure~\ref{fig:densitycut}.

\begin{figure}
    \includegraphics[width=\columnwidth]{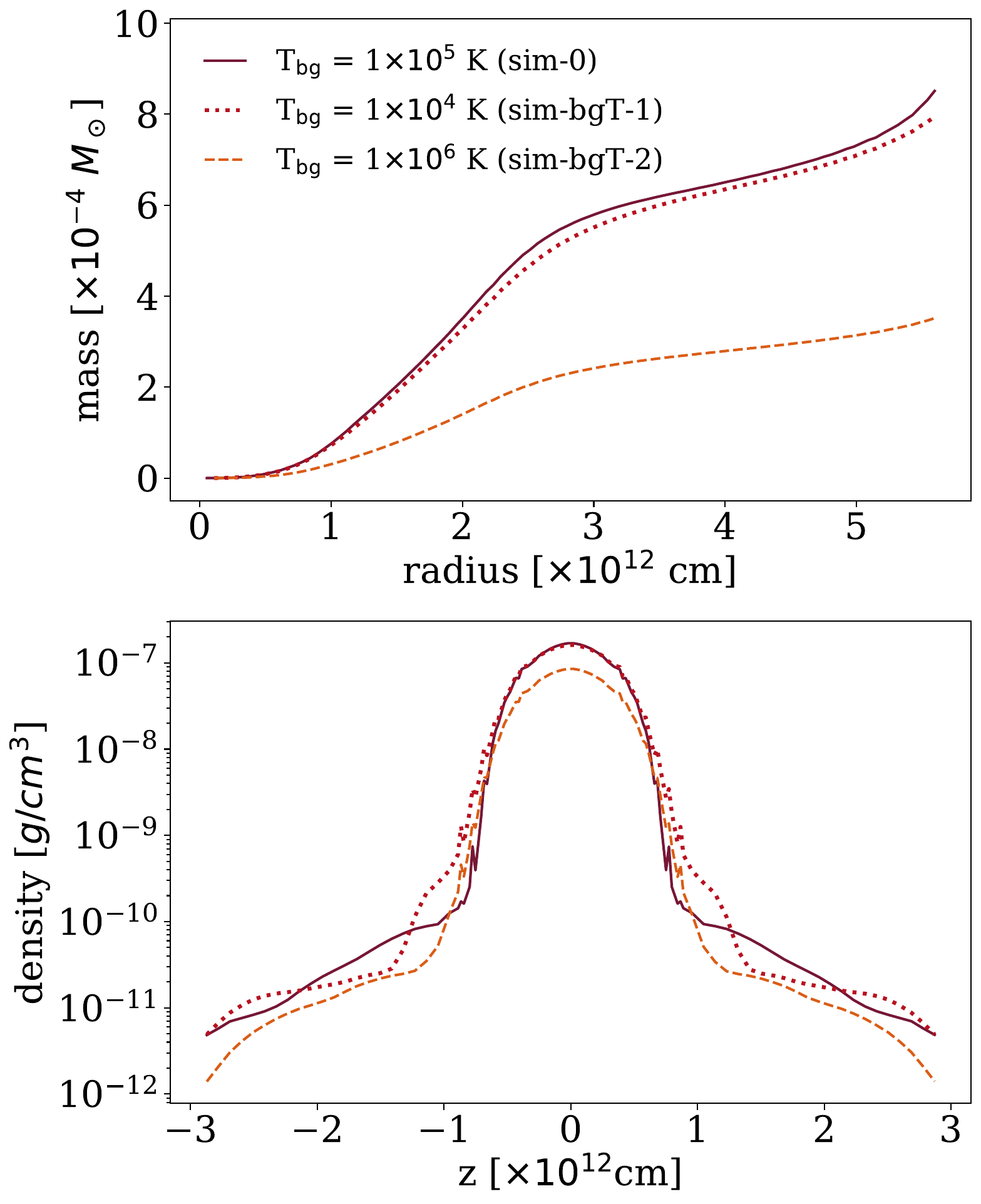}
    \caption{\referee{The cumulative mass as a function of radius (top panel) and the vertical density profile (bottom panel) for models with different background temperatures at t~=~21~yr. Solid purple line: T$_{\rm bg}$~=~$10^{5}$~cm~s$^{-1}$ (model sim-0), dotted red line: T$_{\rm bg}$~=~$10^{4}$~cm~s$^{-1}$ (sim-bgT-1), dashed orange line: T$_{\rm bg}$~=~$10^{6}$~cm~s$^{-1}$ (sim-bgT-2).
    }}
    \label{fig:mediumdiff}
\end{figure}

We repeat the simulation using the same background density but decreasing and increasing the background temperature to $1\times10^{4}$~K (sim-bgT-1) and $1\times10^{6}$~K (sim-bgT-2), respectively (note that sim-0 has a temperature of $1\times10^5$~K). The adiabatic index is $\gamma = 1.1$ in all simulations (see Section~\ref{sec:makita} for a discussion about the adiabatic index). In Figure~\ref{fig:mediumdiff}, we show the total cumulative mass as a function of radius (top panel) and the density profile of the disk along the z-axis (bottom panel) at the end of the simulation. The vertical density profile of the disk is similar for all three simulations, but the highest temperature (and pressure) simulation results in approximately half of the mass in the disk at 21 years. 

Increasing the background temperature by an order of magnitude increases the pressure of the background such that it is higher than the pressure inside the nozzle (see fifth column in Table~\ref{tab:sims}), reducing the amount of mass that can flow out from the nozzle. In the case of sim-bgT-2 only half of the injected material manage to leave the nozzle compared to the reference simulation. 

\subsection{Convergence tests}
\label{sec:convergence}

We finally test the sensitivity of the simulation to spatial and temporal resolution. Our comparison models have the same parameters as sim-0 (see Table~\ref{tab:sims}) with ($72 \times\ 96 \times\ 32$) cells at the coarsest level; we also preserved the size of the computational box ($-d_{\rm L_1}\leq x \leq 1.25d_{\rm L_1}$, $-1.5 d_{\rm L_1} \leq y \leq 1.5 d_{\rm L_1}$ and $-0.5 d_{\rm L_1} \leq z \leq 0.5 d_{\rm L_1}$, where $d_{\rm L_1}=83$~\rsun). We repeated the simulation with increasing levels of refinements: 2 levels (as the low-resolution simulation), 3, 4, and 5 levels (high-resolution simulation). The inner boundary around the companion star has a constant radius of R$_{\rm{in}}~=1.3$~\rsun) in all four simulations, and the size of the nozzle is also not resolution dependent. Refinement occurs primarily in the higher density regions close to the accretor, as concentric circles on the orbital plane, and around the nozzle area; this high-resolution refinement expands along the line that connects the nozzle with the inner boundary on the XZ plane. 

\begin{table}
    \centering
    \begin{tabular}{lcccc}
    \hline
    \hline
         Resolution  &  R$_{\rm disk}$ 
         &  M$_{\rm disk}$  &  H$_{\rm disk}$ & $\dot{\rm M}_{\rm acc}$ \\
         (\# levels) &  (\rsun) 
         &  (\msun) &  (\rsun) & (\msun/yr)\\
    \hline
         2 &  37 &  2.68 $\times\ 10^{-3}$ & 7.2 & 6.5$\times 10^{-3}$ \\
         3 (sim-0)&  39 &  5.46 $\times\ 10^{-3}$ & 4.9 & 4.1$\times 10^{-3}$\\
         4 &  41 &  5.81 $\times\ 10^{-3}$ & 4.5 & 4.2$\times 10^{-3}$\\
         5 &  42 &  6.15 $\times\ 10^{-3}$ & 4.6 & 3.8$\times 10^{-3}$ \\
    \hline
    \end{tabular}
     \caption{Disk properties for the Reference simulation (sim-0), at 21~yr, with different resolutions.}
    \label{tab:convergence}
\end{table}

\begin{figure}
    \centering
    \includegraphics[width=\columnwidth]{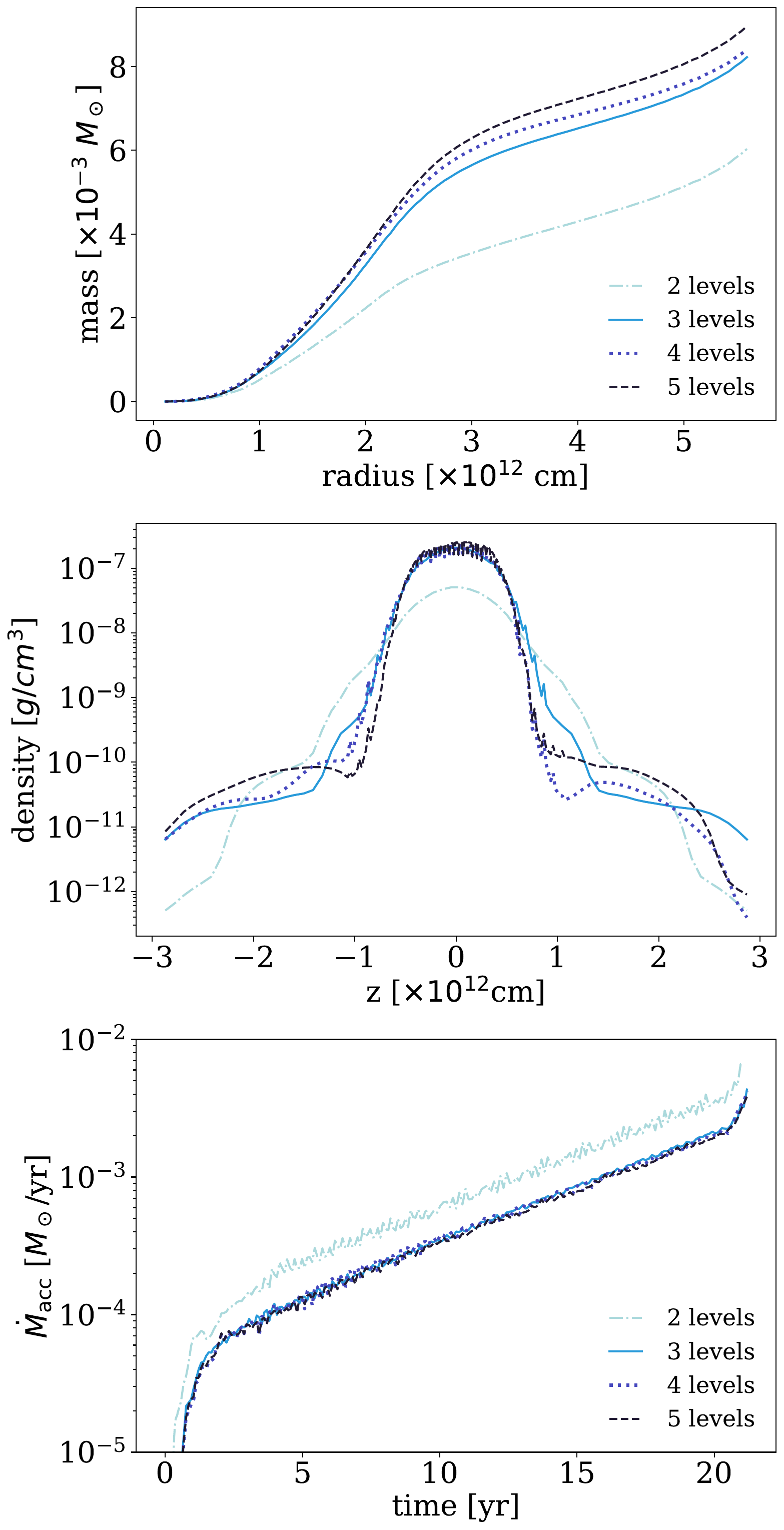}
    \caption{\referee{
        The cumulative mass as a function of radius (top panel), the vertical density profile (middle panel) of the accretion disk at t~=~21 yr for different resolutions. And the mass accretion rate onto the companion (bottom panel) as function of time for different resolutions. sim-0 has 3 levels of refinement.
}}
    \label{fig:convergence}
\end{figure}

In Figure~\ref{fig:convergence}, we show the convergent behaviour of the disk mass and scale height (top and middle panels). We measure the properties of the disk at the end of the simulation (21~yr; see Table~\ref{tab:convergence}). The results show that the disk properties converge across different resolution levels. For the highest resolution (5 levels) simulation, the disk mass shows a 12\% discrepancy compared to the Reference simulation (sim-0; 3 levels of refinement), the disk radius and scale height as measured using the criteria established in Section~\ref{sec:isadisk} are consistent across the three highest resolution simulations.

\referee{We test the convergence of the mass accretion rate onto the neutron star as we have done for other quantities. In the bottom panel of Figure~\ref{fig:convergence}, it is clear that the accretion rate converges. At the highest resolution (5 levels of refinement), the total accreted mass is 7\% smaller than for the reference simulation (3 levels of refinement).}

\subsection{Simulation dimensionality and adiabatic index}
\label{sec:makita}

\referee{The equation of state, given by $P=(\gamma-1)\rho e$, relates the gas pressure $P$ to the internal specific energy $e$ and adiabatic index $\gamma$.   In the absence of dissipation,  a fluid parcel's internal energy changes only due to the $P\,\mathrm{d}V$ work that it does on the surrounding fluid or vice-versa, yielding the adiabatic relationship $P \rho^{-\gamma}  = \mathrm{constant}$.    A higher adiabatic index, implies that a higher fraction of the internal energy is in the kinetic energy of the particles and manifests as pressure, resulting in lower gas compressibility. On the other hand, with a lower adiabatic index the majority of the work done during compression is ``hidden'' in the internal excitation of the particles, so the pressure does not increase as significantly.   The adopted adiabatic index also plays an analogous role in shock-heating, in which bulk kinetic energy is converted to internal energy.
}

The adoption of an isothermal gas was proposed by \citet{lubow1975gas}. \citet{Armitage2000} then showed, using an adiabatic index of 4/3 in their 2D simulations of a mass transfer into the Roche lobe of a neutron star, that an accretion disk could form. Subsequently, \citet{makita2000two}, \citet{macleod2017common}, and \citet{Murguia-Berthier2017AccretionFormation} showed, using 3D simulations of mass transfer through a Roche lobe with various system parameters, that the maximum adiabatic index that would allow a disk to form was $\gamma \leq 1.2$, for simulations without radiation or cooling function implemented.

\begin{figure}
    \centering
    \includegraphics[width=\columnwidth]{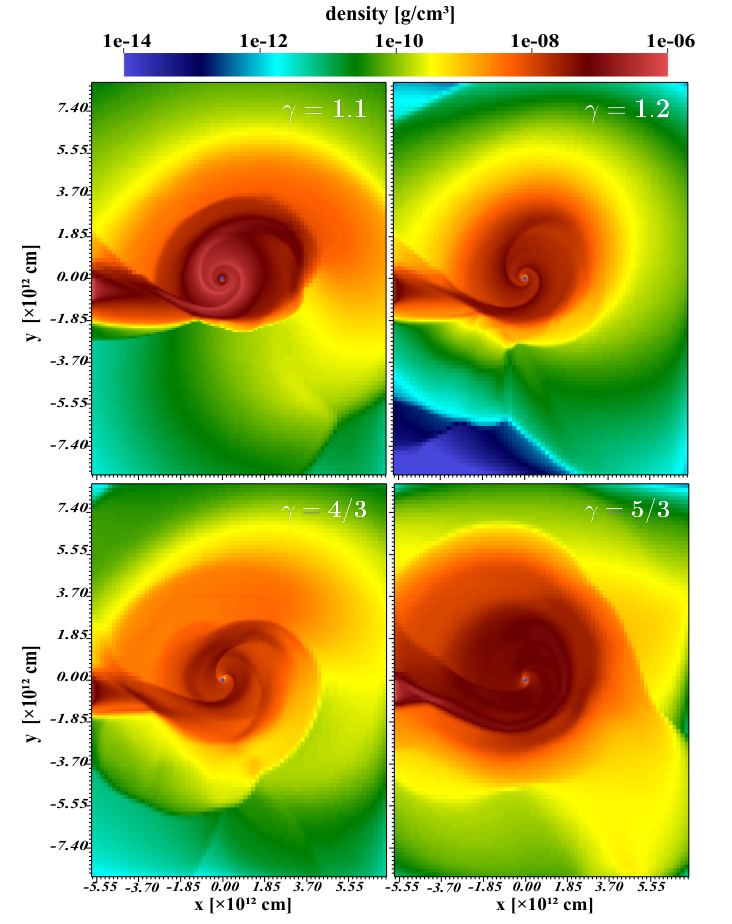}
    \caption{Density in the orbital plane for 2D simulations with different adiabatic indexes. Adiabatic index $\gamma= 1.1$ is on the upper-left panel (a 2D version of the 3D sim-0), $\gamma= 1.2$ in the upper-right panel, $\gamma= 4/3$ in the lower-left panel, and $\gamma= 5/3$ in the lower-right panel). All simulations are plotted at t~=~21~yr.}
    \label{fig:appmakita2d}
\end{figure}

We tested these claims by first performing a 2D comparison simulation of the reference simulation (sim-0, Table~\ref{tab:sims}), and three additional 2D simulations with larger adiabatic indices and then carrying out two 3D simulations with higher adiabatic indices than sim-0. \referee{For the 2D simulations, we  assume the gas is restricted to the orbital plane. We drop the z-coordinate from the grid, $\Delta z = 1$ for all cells, assuming symmetry along the z-axis, and the gas can not be deflected in the z-direction. The code solved the two-dimensional Euler equations for inviscid gas with external forces $\textbf{f}~=~\left(f_{\rm x},f_{\rm y}\right)$. The simulations have the same physical parameters as the reference simulation (sim-0).}

In Figure~\ref{fig:appmakita2d}, we present density slices of the four, 2D simulations with $\gamma=1.1$ (top left panel, a 2D version of sim-0), $1.2$ (top right panel), $4/3$ (bottom left panel) and $5/3$ (bottom right panel), after $21$~yr. In all simulations, the injected material revolves around the companion and forms a high density disk structure. 

\begin{figure}
    \centering
    \includegraphics[width=\columnwidth]{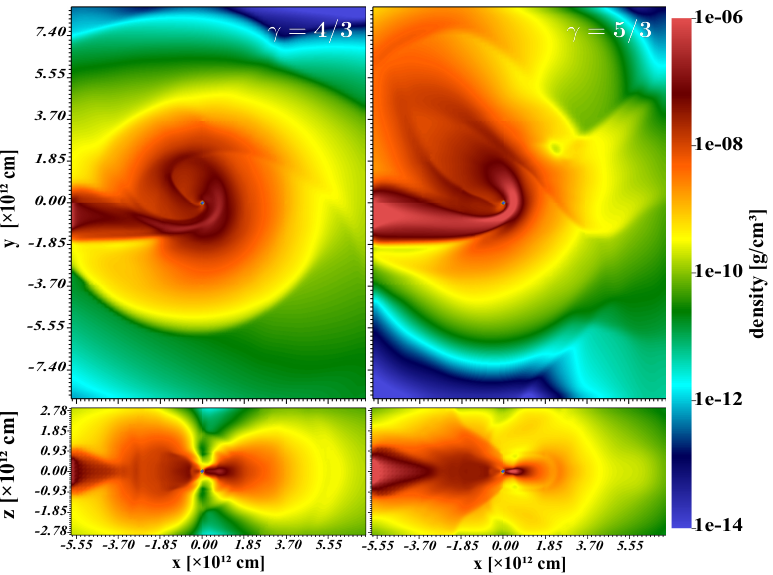}
    \caption{Density slices in the orbital plane (upper panels) and perpendicular plane (lower panels) for 3D simulations with different adiabatic indexes ($\gamma= 4/3$ left, and $\gamma= 5/3$ right). The simulations are plotted at t~=~21~yr, and can be compared with last panel of Fig.~\ref{fig:densitycut}).
}
    \label{fig:appmakita3d}
\end{figure}

In Figure~\ref{fig:appmakita3d} we show density slices of two 3D simulations performed with adiabatic indexes $\gamma=4/3$ and $\gamma = 5/3$ to investigate the reluctance to form a disk for higher values of $\gamma$ observed by other authors. This figure should be compared with sim-0 ($\gamma = 1.1$) shown in the last column of Figure~\ref{fig:densitycut}. For $\gamma = 4/3$, only slightly higher than the value used in sim-0, some of the material gets deflected around the companion, which may prevent the formation of a stable disk. For $\gamma=5/3$, the pressure gradient dominates over the gravitational force of the companion so that a fraction of the material is deflected away from the centre and may inhibit the formation of a disk. 

\begin{figure}
    \centering
    \includegraphics[width=\columnwidth]{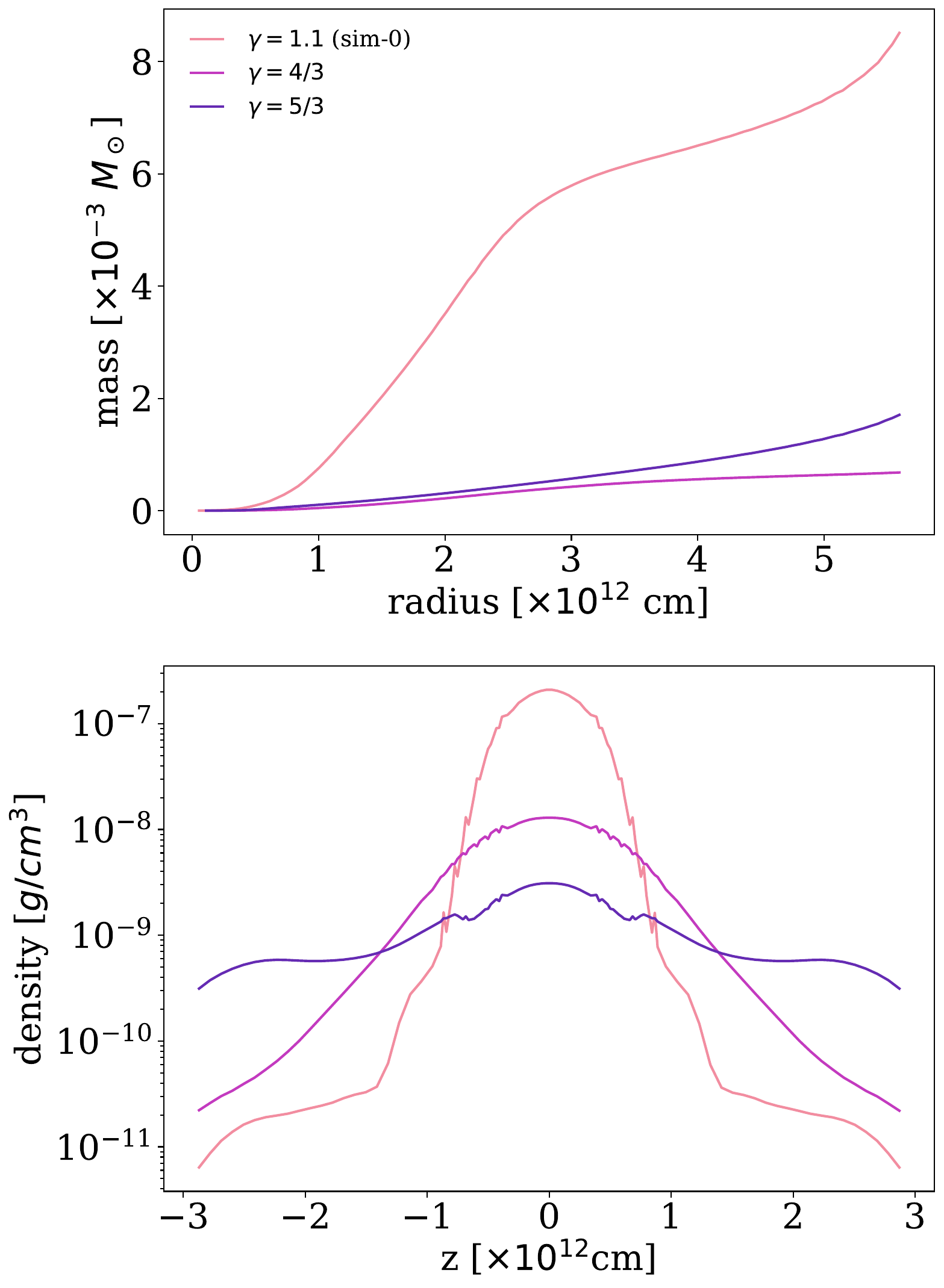}
    \caption{The cumulative mass as a function of radius (top panel) and vertical density profile (bottom panel) for 3D models with different adiabatic index. The simulations are plotted at t~=~21~yr. Density slices of these models are presented in the last column of Figure~\ref{fig:densitycut} and in  Fig.~\ref{fig:appmakita3d}.
}
    \label{fig:analysismakita}
\end{figure}

We quantify the difference between 3D simulations with different  adiabatic indices (see last row of Table~\ref{tab:results}). The top panel of Figure~\ref{fig:analysismakita} shows the cumulative mass as a function of radius, from which we measure the radius and mass of the disk-like structure in each simulation. Even though the simulations have the same mass transfer rates as the reference simulation, the high-density structure formed on the simulation with an adiabatic index $\gamma~=~4/3$ is five times smaller (15.2~\rsun\, or 1.06$\times 10^{12}$~cm) and two orders of magnitude less massive (5.37$\times 10^{-5}$~M$_\odot$). In the case of the simulation with $\gamma~=~5/3$ a radius and a mass value cannot be determined from the cumulative mass plot.

In the bottom panels of Figure~\ref{fig:analysismakita}, we compare the density profile along the z-axis of the simulations with different thermal properties. It is clear that when the adiabatic index is closer to the isothermal value the vertical structure is more defined, as expected from the bottom panels of Figure~\ref{fig:appmakita3d}, while for the higher adiabatic indices there is only a very marginal equatorial compression.

These results are consistent with those of \cite{makita2000two}, \cite{macleod2015asymmetric}, and \cite{Murguia-Berthier2017AccretionFormation}, who showed that disk formation depends primarily on its thermal properties: a disk forms only with an adiabatic index lower than $\gamma\leq 1.2$ when the gas is cooler and has more compressibility. Although, in the case of $\gamma=4/3$ a thick disk seems to be formed in our simulations.

\section{Discussion and conclusion}
\label{sec:Discussion}

The goal of this paper was to study the formation of an accretion disk around a neutron star (1.4~\msun) due to unstable mass transfer from an intermediate mass red supergiant (7~\msun) through Roche lobe overflow (RLOF). This phase likely immediately precedes a common envelope (CE) in-spiral with concomitant accretion onto the companion and possibly the formation of a jet that may affect the pre-in-spiral, as well as the in-spiral phases. Such feedback may lead to a different outcome than what has been modelled thus far by 3D hydrodynamic simulations without feedback  \citep[e.g.,][]{Lau2022} \referee{or by 3D hydrodynamic simulations with a simplified version of feedback \citep[e.g.,][]{Hillel2022}.}

By necessity, with an explicit 3D simulation, we only model a short time (21 years) of the evolution of the RLOF phase predicted to last 30\,000 years by a 1D, implicit model. By carefully choosing to model in 3D the last phases of the mass-transfer before the putative CE in-spiral, we show that the 3D disk mass is likely  only a factor of a few smaller than it might be if the entire phase were modelled, and very similar in radius.

We show that the accretion disk in our system grows to a mass of $\sim 5 \times 10^{-3}$~M$_\odot$, a radius of $\sim 40~$R$_\odot$, and a scale height of $\sim 5~$R$_\odot$, just before it presumably goes into the CE in-spiral phase. This disk has approximately Keplerian rotation near the inner boundary, while the outer regions have a rotation velocity slightly slower than Keplerian due to pressure support. The temperature profile between the inner boundary and the disk's outer radius can be fit with an exponential law with index $-1.1$ steeper than $-0.75$ predicted for accretion disks. An immediate improvement to understanding the disk's temperature (and structure) would be to include explicit cooling. At the one significant figure level, these results are resilient with respect to various physical and numerical choices. The results are well converged with respect to spatial and temporal resolution. 

The accretion rate through the inner boundary that surrounds the companion reaches $4\times 10^{-3}$~\mdot, at the end of the simulation, which is consistent with the accretion rate predicted using a \citet{Shakura1973} formalism with $\alpha=0.1$. \referee{We show that the high accretion rate is driven by turbulent gas motion and possibly shocks, rather than by numerical viscosity. The influence of magnetic fields on the accretion rates cannot be measured in our non-magnetic code, but it remains a possibility in nature. While disk magnetic fields at the inner boundary of the disk should be Gauss-level, such magnetic field strength could be amplified rapidly by the magnetic-rotational instability \citep{Balbus1991} -- within a couple of orbital periods and their effect on disk accretion could therefore be substantial.} 

The Eddington limit for mass accretion onto a neutron star of 1.4~\msun\, is on the order of $10^{-8}$~\mdot. X-ray binaries are known to accrete at rates that can be 100 times the Eddington rate. That said the rate modelled for systems accreting in an unstable mass transfer regime can be larger than that \citep[e.g., 1000 times, see][]{Dickson2024}. Extreme mass accretion rates predicted in the last phases of mass transfer before the CE in-spiral are likely far more complex physical phenomena, with likely extreme X-ray feedback. \cite{mm2022} explained the formation of high mass X-ray binaries, with a black hole as a companion, with hyper accretion rates ($\sim 1 \times 10^{8}~\dot{M}_{\rm Edd}$) before the CE phase.

\referee{Also, the measured maximum accretion rate of $4\times 10^{-4}$~\mdot\ could result in a jet mass loss rate of $4\times 10^{-5}$~\mdot. An escape velocity at the inner boundary of 644~km~s$^{-1}$, would result in a mechanical luminosity of $\sim3\times 10^3$~L$_\odot$. On the other hand, if the jet is launched close to the surface of the neutron star (which is not resolved in our numerical simulations), the escape velocity would be $\sim 0.3$~c, and the mechanical luminosity would be $\sim 10^8$~L$_\odot$.} Understanding the potential for jet formation from disks such as this one would be the next critical step because it may impact the early CE, even if the accretion disk and jet do not survive entering the envelope \citep{Murguia-Berthier2017AccretionFormation}. \referee{Another possibility in this case, could be that the companion jets can prevent the companion from entering the CE. Instead, the companion undergoes ``grazing envelop evolution", where the companion slowly enters the donor's envelope while its jets help unbind the envelope \citep[][]{Shiber2024}. } Further investigation on the survival of the accretion disk during the CE phase for this system is intended on a follow up study.

\paragraph{Acknowledgements}
We acknowledge the computing time granted by DGTIC UNAM on the supercomputer Miztli (projects LANCAD-UNAM-DGTIC-281 and LANCAD-UNAM-DGTIC-321). AJG and FDC acknowledge support from the UNAM-PAPIIT grant IN113424. AJG also acknowledges funding support from Macquarie University
through the International Macquarie University Research Excellence
Scholarship (‘iMQRES’) and the Postgraduate Research Fund ('PGRF').

\paragraph{Data Availability Statement}
The data underlying this article will be shared on reasonable request to the corresponding author.

\printendnotes

\bibliography{example}

\begin{thebibliography}{}
\expandafter\ifx\csname natexlab\endcsname\relax\def\natexlab#1{#1}\fi

\bibitem[{Abbott {et~al.}(2016)Abbott, Abbott, Abbott, Abernathy, Acernese, Ackley, Adams, Adams, Addesso, Adhikari, {et~al.}}]{abbott2016gw151226}
Abbott, B.~P., Abbott, R., Abbott, T., {et~al.} 2016, Physical review letters, 116, 241103

\bibitem[{{Armitage} \& {Livio}(2000)}]{Armitage2000}
{Armitage}, P.~J., \& {Livio}, M. 2000, \apj, 532, 540

\bibitem[{{Balbus} \& {Hawley}(1991)}]{Balbus1991}
{Balbus}, S.~A., \& {Hawley}, J.~F. 1991, \apj, 376, 214

\bibitem[{{Brown} {et~al.}(2007){Brown}, {Lee}, \& {Moreno M{\'e}ndez}}]{Brown2007GRB}
{Brown}, G.~E., {Lee}, C.-H., \& {Moreno M{\'e}ndez}, E. 2007, \apjl, 671, L41

\bibitem[{Cehula \& Pejcha(2023)}]{Cehula2023}
Cehula, J., \& Pejcha, O. 2023, Monthly Notices of the Royal Astronomical Society, 524, 471

\bibitem[{{Chamandy} {et~al.}(2018){Chamandy}, {Frank}, {Blackman}, {Carroll-Nellenback}, {Liu}, {Tu}, {Nordhaus}, {Chen}, \& {Peng}}]{chamandy2018}
{Chamandy}, L., {Frank}, A., {Blackman}, E.~G., {et~al.} 2018, \mnras, 480, 1898

\bibitem[{Chevalier(2012)}]{chevalier2012common}
Chevalier, R.~A. 2012, The Astrophysical Journal Letters, 752, L2

\bibitem[{De~Colle {et~al.}(2012)De~Colle, Granot, L{\'o}pez-C{\'a}mara, \& Ramirez-Ruiz}]{decolle2012}
De~Colle, F., Granot, J., L{\'o}pez-C{\'a}mara, D., \& Ramirez-Ruiz, E. 2012, The Astrophysical Journal, 746, 122

\bibitem[{{Dickson}(2024)}]{Dickson2024}
{Dickson}, D. 2024, \apj, 975, 130

\bibitem[{{Eggleton}(1983)}]{eggleton1983}
{Eggleton}, P.~P. 1983, \apj, 268, 368

\bibitem[{Fryer \& Woosley(1998)}]{fryer1998helium}
Fryer, C., \& Woosley, S. 1998, The Astrophysical Journal, 502, L9

\bibitem[{{Hayashi} {et~al.}(2021){Hayashi}, {Kawaguchi}, {Kiuchi}, {Kyutoku}, \& {Shibata}}]{2021Hayashi}
{Hayashi}, K., {Kawaguchi}, K., {Kiuchi}, K., {Kyutoku}, K., \& {Shibata}, M. 2021, \prd, 103, 043007

\bibitem[{{Hillel} {et~al.}(2022){Hillel}, {Schreier}, \& {Soker}}]{Hillel2022}
{Hillel}, S., {Schreier}, R., \& {Soker}, N. 2022, \mnras, 514, 3212

\bibitem[{{Iben} \& {Tutukov}(1984)}]{ibenandttukov1984}
{Iben}, I., J., \& {Tutukov}, A.~V. 1984, \apjs, 54, 335

\bibitem[{Ivanova {et~al.}(2013)Ivanova, Justham, Chen, De~Marco, Fryer, Gaburov, Ge, Glebbeek, Han, Li, {et~al.}}]{ivanova2013common}
Ivanova, N., Justham, S., Chen, X., {et~al.} 2013, The Astronomy and Astrophysics Review, 21, 1

\bibitem[{Jackson {et~al.}(2017)Jackson, Arras, Penev, Peacock, \& Marchant}]{Jackson2017AStars}
Jackson, B., Arras, P., Penev, K., Peacock, S., \& Marchant, P. 2017, The Astrophysical Journal, 835, 145

\bibitem[{{Lau} {et~al.}(2022){Lau}, {Hirai}, {Gonz{\'a}lez-Bol{\'\i}var}, {Price}, {De Marco}, \& {Mandel}}]{Lau2022}
{Lau}, M. Y.~M., {Hirai}, R., {Gonz{\'a}lez-Bol{\'\i}var}, M., {et~al.} 2022, \mnras, 512, 5462

\bibitem[{{L{\'o}pez-C{\'a}mara} {et~al.}(2019){L{\'o}pez-C{\'a}mara}, {De Colle}, \& {Moreno M{\'e}ndez}}]{lopez2019}
{L{\'o}pez-C{\'a}mara}, D., {De Colle}, F., \& {Moreno M{\'e}ndez}, E. 2019, \mnras, 482, 3646

\bibitem[{{L{\'o}pez-C{\'a}mara} {et~al.}(2022){L{\'o}pez-C{\'a}mara}, {De Colle}, {Moreno M{\'e}ndez}, {Shiber}, \& {Iaconi}}]{LopezCamara2022}
{L{\'o}pez-C{\'a}mara}, D., {De Colle}, F., {Moreno M{\'e}ndez}, E., {Shiber}, S., \& {Iaconi}, R. 2022, \mnras, 513, 3634

\bibitem[{L{\'o}pez-C{\'a}mara {et~al.}(2020)L{\'o}pez-C{\'a}mara, Moreno~M{\'e}ndez, \& De~Colle}]{lopez2020disc}
L{\'o}pez-C{\'a}mara, D., Moreno~M{\'e}ndez, E., \& De~Colle, F. 2020, Monthly Notices of the Royal Astronomical Society, 497, 2057

\bibitem[{Lubow \& Shu(1975)}]{lubow1975gas}
Lubow, S.~H., \& Shu, F.~H. 1975, The Astrophysical Journal, 198, 383

\bibitem[{MacLeod {et~al.}(2017)MacLeod, Antoni, Murguia-Berthier, Macias, \& Ramirez-Ruiz}]{macleod2017common}
MacLeod, M., Antoni, A., Murguia-Berthier, A., Macias, P., \& Ramirez-Ruiz, E. 2017, The Astrophysical Journal, 838, 56

\bibitem[{MacLeod \& Ramirez-Ruiz(2015)}]{macleod2015asymmetric}
MacLeod, M., \& Ramirez-Ruiz, E. 2015, The Astrophysical Journal, 803, 41

\bibitem[{Makita {et~al.}(2000)Makita, Miyawaki, \& Matsuda}]{makita2000two}
Makita, M., Miyawaki, K., \& Matsuda, T. 2000, Monthly Notices of the Royal Astronomical Society, 316, 906

\bibitem[{Mineshige {et~al.}(1994)Mineshige, Honma, Hirano, Kitamoto, Yamada, \& Fukue}]{Mineshige94}
Mineshige, S., Honma, F., Hirano, A., {et~al.} 1994, in Theory of Accretion Disks --- 2, ed. W.~J. Duschl, J.~Frank, F.~Meyer, E.~Meyer-Hofmeister, \& W.~M. Tscharnuter (Dordrecht: Springer Netherlands), 187--193

\bibitem[{{Moe} \& {Di Stefano}(2017)}]{moedisteffano2017}
{Moe}, M., \& {Di Stefano}, R. 2017, \apjs, 230, 15

\bibitem[{{Moreno M{\'e}ndez}(2022)}]{mm2022}
{Moreno M{\'e}ndez}, E. 2022, arXiv e-prints, arXiv:2207.14765

\bibitem[{{Moreno M{\'e}ndez} {et~al.}(2017){Moreno M{\'e}ndez}, {L{\'o}pez-C{\'a}mara}, \& {De Colle}}]{moreno2017}
{Moreno M{\'e}ndez}, E., {L{\'o}pez-C{\'a}mara}, D., \& {De Colle}, F. 2017, \mnras, 470, 2929

\bibitem[{Murguia-Berthier {et~al.}(2017)Murguia-Berthier, MacLeod, Ramirez-Ruiz, Antoni, \& Macias}]{Murguia-Berthier2017AccretionFormation}
Murguia-Berthier, A., MacLeod, M., Ramirez-Ruiz, E., Antoni, A., \& Macias, P. 2017, The Astrophysical Journal, 845, 173

\bibitem[{{Paxton} {et~al.}(2011){Paxton}, {Bildsten}, {Dotter}, {Herwig}, {Lesaffre}, \& {Timmes}}]{paxton2011}
{Paxton}, B., {Bildsten}, L., {Dotter}, A., {et~al.} 2011, \apjs, 192, 3

\bibitem[{{Paxton} {et~al.}(2013){Paxton}, {Cantiello}, {Arras}, {Bildsten}, {Brown}, {Dotter}, {Mankovich}, {Montgomery}, {Stello}, {Timmes}, \& {Townsend}}]{paxton2013}
{Paxton}, B., {Cantiello}, M., {Arras}, P., {et~al.} 2013, \apjs, 208, 4

\bibitem[{{Paxton} {et~al.}(2015){Paxton}, {Marchant}, {Schwab}, {Bauer}, {Bildsten}, {Cantiello}, {Dessart}, {Farmer}, {Hu}, {Langer}, {Townsend}, {Townsley}, \& {Timmes}}]{paxton2015}
{Paxton}, B., {Marchant}, P., {Schwab}, J., {et~al.} 2015, \apjs, 220, 15

\bibitem[{{Paxton} {et~al.}(2018){Paxton}, {Schwab}, {Bauer}, {Bildsten}, {Blinnikov}, {Duffell}, {Farmer}, {Goldberg}, {Marchant}, {Sorokina}, {Thoul}, {Townsend}, \& {Timmes}}]{paxton2018}
{Paxton}, B., {Schwab}, J., {Bauer}, E.~B., {et~al.} 2018, \apjs, 234, 34

\bibitem[{{Paxton} {et~al.}(2019){Paxton}, {Smolec}, {Schwab}, {Gautschy}, {Bildsten}, {Cantiello}, {Dotter}, {Farmer}, {Goldberg}, {Jermyn}, {Kanbur}, {Marchant}, {Thoul}, {Townsend}, {Wolf}, {Zhang}, \& {Timmes}}]{paxton2019}
{Paxton}, B., {Smolec}, R., {Schwab}, J., {et~al.} 2019, \apjs, 243, 10

\bibitem[{{Pjanka} \& {Stone}(2020)}]{2020ApJ...904...90P}
{Pjanka}, P., \& {Stone}, J.~M. 2020, \apj, 904, 90

\bibitem[{{Pringle} \& {King}(2014)}]{Pringle2014}
{Pringle}, J.~E., \& {King}, A. 2014, {Astrophysical Flows} (Cambridge University Press)

\bibitem[{{Ramachandran} {et~al.}(2022){Ramachandran}, {Oskinova}, {Hamann}, {Sander}, {Todt}, {Pauli}, {Shenar}, {Torrej{\'o}n}, {Postnov}, {Blondin}, {Bozzo}, {Hainich}, \& {Massa}}]{Ramachandra22}
{Ramachandran}, V., {Oskinova}, L.~M., {Hamann}, W.~R., {et~al.} 2022, \aap, 667, A77

\bibitem[{Ramirez-Ruiz \& Lee(2009)}]{ramirez2009maybe}
Ramirez-Ruiz, E., \& Lee, W. 2009, Nature, 460, 1091

\bibitem[{{Rea} \& {De Grandis}(2025)}]{Rea2025}
{Rea}, N., \& {De Grandis}, D. 2025, arXiv e-prints, arXiv:2503.04442

\bibitem[{{Ritter}(1988)}]{Ritter1988}
{Ritter}, H. 1988, \aap, 202, 93

\bibitem[{Savonije(1977)}]{Savonije77}
Savonije, G. 1977, Astronomy and Astrophysics, 62, 317

\bibitem[{{Shakura} \& {Sunyaev}(1973)}]{Shakura1973}
{Shakura}, N.~I., \& {Sunyaev}, R.~A. 1973, \aap, 24, 337

\bibitem[{Shakura \& Sunyaev(1973)}]{shakura1973black}
Shakura, N.~I., \& Sunyaev, R.~A. 1973, Astronomy and Astrophysics, 24, 337

\bibitem[{{Shiber} \& {Iaconi}(2024)}]{Shiber2024}
{Shiber}, S., \& {Iaconi}, R. 2024, \mnras, 532, 692

\bibitem[{{Shiber} {et~al.}(2019){Shiber}, {Iaconi}, {De Marco}, \& {Soker}}]{Shiber2019}
{Shiber}, S., {Iaconi}, R., {De Marco}, O., \& {Soker}, N. 2019, \mnras, 488, 5615

\bibitem[{{Tauris} {et~al.}(2017){Tauris}, {Kramer}, {Freire}, {Wex}, {Janka}, {Langer}, {Podsiadlowski}, {Bozzo}, {Chaty}, {Kruckow}, {van den Heuvel}, {Antoniadis}, {Breton}, \& {Champion}}]{tauris2017}
{Tauris}, T.~M., {Kramer}, M., {Freire}, P.~C.~C., {et~al.} 2017, \apj, 846, 170

\bibitem[{{Wardle}(2007)}]{Wardle2007}
{Wardle}, M. 2007, \apss, 311, 35

\bibitem[{Warner(1995)}]{warner_1995}
Warner, B. 1995, Cataclysmic Variable Stars, Cambridge Astrophysics (Cambridge University Press), doi:10.1017/CBO9780511586491

\end{thebibliography}
\appendix
\renewcommand{\thefigure}{A\arabic{figure}} 
\setcounter{figure}{0}
\section{Numerical considerations}

\subsection{Conservation of mass, energy and angular momentum}
\label{sec:conservation}

As we explain in Sec. \ref{sssec:3D}, the Reference simulation has constant mass injection through the nozzle, we have external outflow boundaries and the inner inflow boundary. To test for conservation, we run the reference simulation, for 9.5 years after which we switch off the nozzle, change the outflow boundary condition  into reflective boundaries, and remove the outflow boundary around the companion star in the middle of the domain. We then run the simulation for an additional 11.5 years, during which we evaluate the conservation of mass, angular momentum, and energy.

\begin{figure}
    \centering
    \includegraphics[width=\columnwidth]{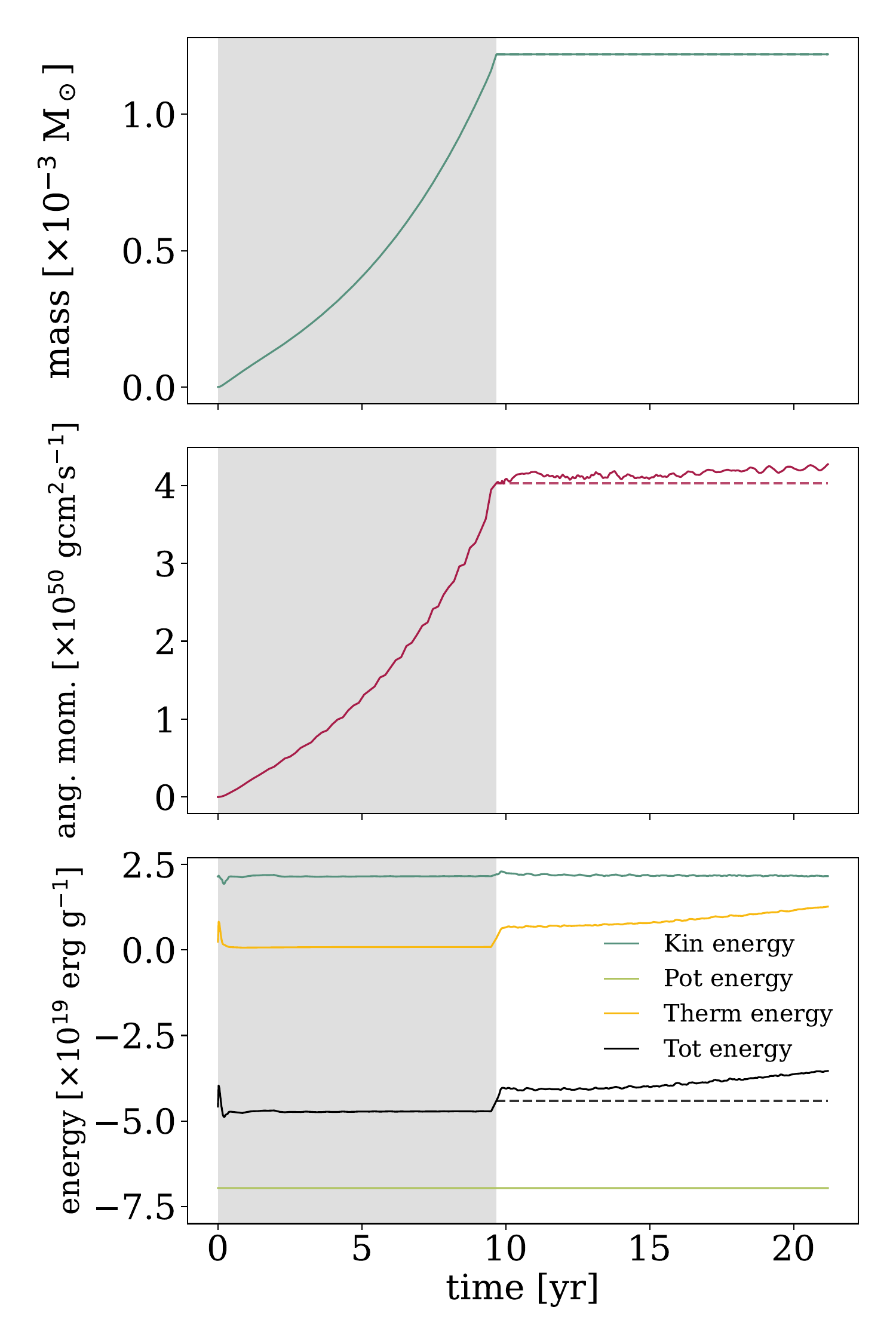}
    \caption{\referee{Total mass (top panel), total angular momentum (middle panel) within the computational domain as a function of time. The bottom panel shows total kinetic energy (Kin energy), potential energy (Pot energy), thermal energy (Therm energy), and total energy (Tot energy) as a function of time. During the grey shadow area, the mass injection is on, and internal inflow boundary is present on the computational box. The dash line in each panel represents the comparative value after the injection is over.}}
    \label{fig:conservation}
\end{figure}

For each time step, we calculate the total mass, angular momentum, and energy in the inertial frame of reference. For this we integrate the mass, angular momentum in each cell at every time step. In the case of the total energy, we take into account the kinetic energy of each cell, the potential energy between the fluid and the companion star, and the thermal energy of each cell. The integrated values of mass, angular momentum and energy are shown in Figure~\ref{fig:conservation}, where the grey area represents the time when the nozzle and mass injection are on, causing the total mass, energy and angular momentum to increase over time. The dashed light line in each panel represents the first value of mass, energy angular momentum after the nozzle is switched off.

As shown in the top panel of Figure~\ref{fig:conservation}, the mass within the reflective boundaries remains constant throughout the simulation, with a maximum variation of $~0.02$~\% at $21$~yr. As expected, due to numerical effects such as numerical viscosity, and the way the conservation of the angular momentum equation is discretised, in the middle panel of Figure~\ref{fig:conservation}, the total angular momentum grows up to $6\%$ over the 10 years the simulation has reflective boundaries. For the measure of the total energy within the computational box (bottom panel in Figure~\ref{fig:conservation}), after turning off the nozzle, there is a decrease in kinetic energy over 0.15~yr. This change is reflected in a step increase in thermal energy of 2.5$\times10^{18}$~erg~s$^{-1}$. The energy has a maximum change of $13\%$ after the mass injection finish and remains constant until the end of the simulation. The extent of the non-conservation of the preceding values justifies the approximations made in the simulations.

\end{document}